# Search for "anomalies" from neutrino and anti-neutrino oscillations at Δm² ≈ 1eV² with muon spectrometers and large LAr–TPC imaging detectors.

## Technical proposal.

### (CERN-SPSC-2012-010 and SPSC-P-347)

## ICARUS Collaboration

M. Antonello[1], D. Bagliani[2], B. Baibussinov[5], H. Bilokon[6], F. Boffelli[8], M. Bonesini[9], E. Calligarich[8], N. Canci[1], S. Centro[4,5], A. Cesana[10], K. Cieslik[11], D. B. Cline[12], A. G. Cocco[14], D. Dequal[4,5], A. Dermenev[16], R. Dolfini[7,8], M. De Gerone[2,3], S. Dussoni[2,3], C. Farnese[4], A. Fava[5], A. Ferrari[17], G. Fiorillo[13,14], G. T. Garvey[18], F. Gatti[2,3], D. Gibin[4,5], S. Gninenko[16], F. Guber[16], A. Guglielmi[5], M. Haranczyk[11], J. Holeczek[19], A. Ivashkin[16], M. Kirsanov[16], J. Kisiel[19], I. Kochanek[19], A. Kurepin[16], J. Łagoda[20], G. Lucchini[9], W. C. Louis[18], S. Mania[19], G. Mannocchi[6], S. Marchini[5], V. Matveev[16], A. Menegolli[7,8], G. Meng[5], G. B. Mills[18], C. Montanari[8], M. Nicoletto[5], S. Otwinowski[12], T. J. Palczewski[20], G. Passardi[17], F. Perfetto[13,14], P. Picchi[6], F. Pietropaolo[5], P. Płoński[21], A. Rappoldi[8], G. L. Raselli[8], M. Rossella[8], C. Rubbia[1,17,a], P. Sala[10], A. Scaramelli[10], E. Segreto[1], D. Stefan[1], J. Stepaniak[20], R. Sulej[20], O. Suvorova[16], M. Terrani[10], D. Tlisov[16], R. G. Van de Water[18], G. Trinchero[6], M. Turcato[5], F. Varanini[4], S. Ventura[5], C. Vignoli[1], H. G. Wang[12], X. Yang[12], A. Zani[8], K. Zaremba[21]

*(a)* Contact Person

## NESSiE Collaboration

M. Benettoni[5], P. Bernardini[26,27], A. Bertolin[5], C. Bozza[31], R. Brugnera[4,5], A. Cecchetti[6], S. Cecchini[25], G. Collazuol[5,6], P. Creti[27], F. Dal Corso[5], I. De Mitri[26,27], G. De Robertis[23], M. De Serio[23], L. Degli Esposti[25], D. Di Ferdinando[25], U. Dore[29,30], S. Dusini[5], P. Fabbricatore[2], C. Fanin[5], R. A. Fini[23], G. Fiore[27], A. Garfagnini[4,5], G. Giacomelli[24,25], R. Giacomelli[25], G. Grella[31], C. Guandalini[25], M. Guerzoni[25], U. Kose[5], G. Laurenti[25], M. Laveder[4,5], I. Lippi[5], F. Loddo[23], A. Longhin[6], P. Loverre[29,30], G. Mancarella[26,27], G. Mandrioli[25], A. Margiotta[24,25], G. Marsella[27,28], N. Mauri[6], E. Medinaceli[4,5], A. Mengucci[6], M. Mezzetto[5], R. Michinelli[25], M. T. Muciaccia[22,23], D. Orecchini[6], A. Paoloni[6], A. Pastore[22,23], L. Patrizii[25], M. Pozzato[24,25], R. Rescigno[31], G. Rosa[29], S. Simone[22,23], M. Sioli[24,25], G. Sirri[25], M. Spurio[24,25], L. Stanco[5,b], S. Stellacci[31], A. Surdo[27], M. Tenti[24,25], V. Togo[25], M. Ventura[6] and M. Zago[5].

*(b)* Contact Person

1. *INFN, LNGS, Assergi (AQ), Italy*
2. *Dipartimento di Fisica, Università di Genova, Genova, Italy*
3. *INFN, Sezione di Genova, 16146 Genova, Italy*
4. *Dipartimento di Fisica, Università di Padova, Padova, Italy*
5. *INFN, Sezione di Padova, 35131 Padova, Italy*
6. *INFN, LNF, 00044 Frascati (Roma), Italy*
7. *Dipartimento di Fisica Nucleare e Teorica, Università di Pavia, 27100 Pavia, Italy*
8. *INFN, Sezione di Pavia, 27100 Pavia, Italy*
9. *INFN, Sezione di Milano Bicocca, Dipartimento di Fisica G. Occhialini, 20126 Milano, Italy*
10. *INFN, Sezione di Milano e Politecnico, 20133 Milano, Italy*
11. *The Henryk Niewodniczanski Institute of Nuclear Physics, Polish Academy of Science, Kraków, Poland*
12. *Department of Physics and Astronomy, University of California, Los Angeles, USA*
13. *Dipartimento di Scienze Fisiche, Università Federico II, 80126 Napoli, Italy*
14. *INFN, Sezione di Napoli, 80126 Napoli, Italy*
16. *INR-RAS, Moscow, Russia*
17. *CERN, Geneva, Switzerland*
18. *Los Alamos National Laboratory, New Mexico, USA*
19. *Institute of Physics, University of Silesia, Katowice, Poland*
20. *National Center for Nuclear Research, Warszawa, Poland*
21. *Institute for Radioelectronics, Warsaw University of Technology, Warsaw, Poland*
22. *Dipartimento di Fisica dell'Università di Bari, 70126 Bari, Italy*
23. *INFN, Sezione di Bari, 70126 Bari, Italy*
24. *Dipartimento di Fisica dell'Università di Bologna, 40127 Bologna, Italy*
25. *INFN, Sezione di Bologna, 40127 Bologna, Italy*
26. *Dipartimento di Fisica dell'Università del Salento, 73100 Lecce, Italy*
27. *INFN, Sezione di Lecce, 73100 Lecce, Italy*
28. *Dipartimento di Ingegneria dell'Innovazione dell'Università del Salento, 73100 Lecce, Italy*
29. *Dipartimento di Fisica dell'Università di Roma "La Sapienza", 00185 Roma, Italy*
30. *INFN, Sezione di Roma I, 00185 Roma, Italy*
31. *Dipartimento di Fisica dell'Università di Salerno and INFN, 84084 Fisciano, Salerno, Italy*

# ABSTRACT

The present proposal describes an experimental search for sterile neutrinos beyond the Standard Model with a new CERN-SPS neutrino beam. The core of the experiment will be the presently operational ICARUS T600 imaging detector, the largest LAr-TPC ever built with a size of about 600 t of imaging mass, now running in the LNGS underground laboratory exposed to the CNGS neutrino beam.

The experiment is based on two identical LAr-TPC's followed by magnetized spectrometers, observing the electron and muon neutrino events at the "Far" and "Near" positions 1600 and 300 m from the proton target, respectively. This project will exploit the ICARUS T600, moved from Gran Sasso to the CERN "Far" position. An additional 1/4 of the T600 detector (T150) will be constructed and located in the "Near" position.

Two spectrometers will be placed downstream of the two LAr-TPC detectors to greatly complement the physics capabilities. Spectrometers will exploit a classical dipole magnetic field with iron slabs, and a new concept air-magnet, to perform charge identification and muon momentum measurements from low energy (< 1 GeV) in a wide energy range over a large transverse area (> 50 m$^2$).

In the two positions, the radial and energy spectra of the $\nu_e$ beam are practically identical. Comparing the two detectors, in absence of oscillations, all cross sections and experimental biases cancel out, and the two experimentally observed event distributions must be identical. Any difference of the event distributions at the locations of the two detectors might be attributed to the possible existence of $\nu$-oscillations, presumably due to additional neutrinos with a mixing angle $\sin^2(2\theta_{new})$ and a larger mass difference $\Delta m^2_{new}$.

The superior quality of the LAr imaging TPC, now widely experimentally demonstrated, and in particular its unique electron–$\pi^0$ discrimination allows full rejection of backgrounds and offers a lossless $\nu_e$ detection capability. The determination of the muon charge with the spectrometers allows the full separation of $\nu_\mu$ from $\bar{\nu}_\mu$ and therefore controlling systematics from muon mis-identification largely at high momenta.

Two main anomalies will be explored with both neutrino and anti-neutrino focused beams. According to the first anomaly some of the $\nu_e (\bar{\nu}_e)$ and/or of the $\nu_\mu (\bar{\nu}_\mu)$ events might be converted into invisible (sterile) components, leading to observation of oscillatory, distance dependent disappearance rates. In a second anomaly (following LSND and MiniBooNE observations) some distance dependent $\nu_\mu \rightarrow \nu_e$ oscillations may be observed as $\nu_e$ excess, especially in the antineutrino channel. The disentangling of $\nu_\mu$ from $\bar{\nu}_\mu$ will allow to exploit the interplay of the different possible oscillation scenario, as well as the interplay between disappearance and appearance of different neutrino states and flavors. Moreover the NC/CC ratio will provide a sterile neutrino oscillation signal by itself and it will beautifully complement the normalization and the systematics studies. A total LAr mass of 760 + 200 ton, completed by the two magnetized spectrometers, and a reasonable utilization of a new CERN-SPS neutrino beam line will offer remarkable discovery potentialities, collecting a very large number of unbiased events both in the neutrino and antineutrino channels, largely adequate to definitely settle the origin of the $\nu$-related anomalies.

# Table of contents.

# 1  Introduction.

The present Technical Proposal is the follow-up of previous documents sent to SPS Committee by the ICARUS and NESSiE Collaborations:

- a Memorandum from the ICARUS collaboration (SPSC-M-773) of March 9, 2011 [1];

- a Proposal entitled "*A comprehensive search for <<anomalies>> from neutrino and anti-neutrino oscillations at large mass differences ($\Delta m^2 \approx 1eV^2$) with two LAr–TPC imaging detectors at different distances from the CERN-PS*" describing a search for anomalies from neutrino and anti-neutrino oscillations with two LAr–TPC imaging detectors at different distances from the CERN-PS (SPSC-P-345) of October 14th, 2011 [2];

- a Proposal entitled "*Prospect for Charge Current Neutrino Interactions Measurements at the CERN-PS*" with two magnetic spectrometers for measuring CC neutrino interactions (SPSC-P-343) of October 11th, 2011 [3].

According to the recommendation of the SPS-C, the two proposals are hereby merged into a common Technical Proposal.

Both proposals have been originally designed for a neutrino beam produced from the CERN-PS. More recently the CERN Management has advanced the alternative of using instead a new neutrino beam from the CERN-SPS in the North Area (see Figure 1). The wider space available in the North Area allows a longer distance of up to 1.6 km to be exploited by the Far detector, instead of the 850 m in the PS option. The distance for the Near detector will also be increased from 127 m to about 300 m. Because of the longer baseline (L), the central value of the neutrino beam energy spectrum is correspondingly doubled from $E_\nu \sim 1$ GeV to $\sim 2$ GeV to exploit same neutrino "$L/E_\nu$" ratio for the experiment. As in the present CNGS facility, due to the SPS fast proton extraction, the neutrino pulse duration will be 10.5 $\mu$s, longer than in the PS case but still adequate for the experiment requirements. The energy of the proton beam for on-axis neutrinos has to be sufficiently low in order to produce high intensity neutrinos at $\sim 2$ GeV and to keep under control the beam related background expected at the Near Detector. A proton energy in the region of 100 GeV is deemed to be an acceptable compromise, which however has to be analyzed in more detail with dedicated FLUKA [4] calculations and recognized as acceptable for SPS operation. It is also assumed that the actual operation of the SPS should provide a sufficient intensity in order to achieve at least the integrated neutrino event rates expected by the former studied PS neutrino beam.

We refer to the above mentioned documents for the physics motivations and the experimental signatures of the various anomalies to be investigated. The present Technical Proposal describes in detail the design and locations of the CERN neutrino detectors, namely the moving and the related modifications of the T600 detector, the construction and assembling of the associated muon spectrometer, as well as the construction of the additional T150 LAr-TPC with the downstream muon spectrometer. The detailed design and development of the new neutrino beam will be performed under the responsibility of the CERN Management.

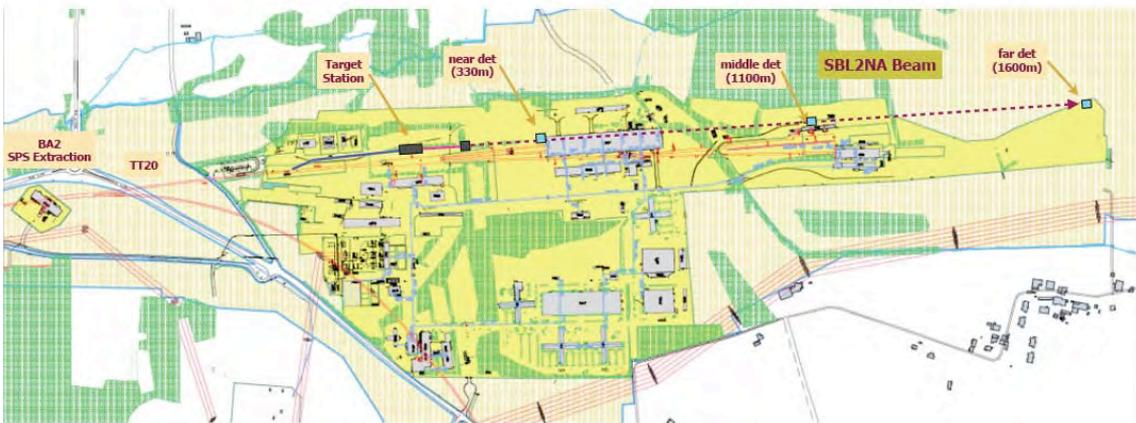

**Figure 1.** The new SPS North Area neutrino beam layout. Main parameters are: primary beam : 100 GeV; fast extracted from SPS; target station next to TCC2, ~11m underground; decay pipe: 100m, 3 m diameter; beam dump : 15m of Fe with graphite core, followed by muon stations; neutrino beam angle: pointing upwards; at -3m in the far detector ~5mrad slope. From I. Efthymiopoulos / CERN.

### 1.1    Beam requirements.

The neutrino beam shall fulfil requirements in terms of intensity and quality that are crucial for the outcome of the experiment. These requirements drive the choice of the primary beam and optics parameters.

As already mentioned, provided the available baseline L, the neutrino beam energy shall be centred around 2 GeV. In principle, this condition can be achieved either with an off-axis configuration, or with an on-axis low energy focusing. However, an off-axis beam on such a short baseline cannot guarantee the similarity of the neutrino beam spectra at the near and far locations, a basic feature for the intrinsic $\nu_e$ component in $\nu_\mu \rightarrow \nu_e$ appearance search. Therefore, only the on-axis beam option is considered in this proposal since it matches the requirements of the experiment.

Beam simulations have been performed assuming a preliminary beam optics where the horn and reflector have a 160 cm diameter, length of 400 and 390 cm, respectively, and are operated at 300 and 150 kA. The target is a 1 m long and 4 mm diameter graphite rod partially inserted into the horn neck. The beam size has been assumed equal to the CNGS one ($\sigma = 0.53$ mm), and a decay path of ~ 100 m is considered. All the target and optics parameters will be subject to further optimization.

According to simulations, the Near detector should be placed at about 300 m from target in order to minimize the oscillation probability over the largest possible $\Delta m^2$ range. The 1600 m position results to be the best choice to maximize the signal to background ratio in the Far detector.

Despite the low-energy focusing, mesons produced in the forward direction will reach the decay tunnel producing both high energy tails in the neutrino spectra and a large amount of energetic muons (Figure 2). These muons have to be absorbed in the beam dump and in the earth before reaching the Near detector. The foreseen 200 m of earth between the end of the decay tunnel and the Near detector should be sufficient to stop muons with energy below 100 GeV. A full simulation is going on to fully assess this point. It is however evident that a much higher energy beam would be unacceptable. For instance, with a 200 GeV proton beam about $10^9$ muons with $E_\mu > 100$ GeV would be produced in a CNGS-like spill of $2 \times 10^{13}$ pot (protons on target) intensity. On the other hand, the operation of the SPS and the transfer line disfavours the use of even lower energy protons. Therefore, the optimal beam energy is around 100 GeV.

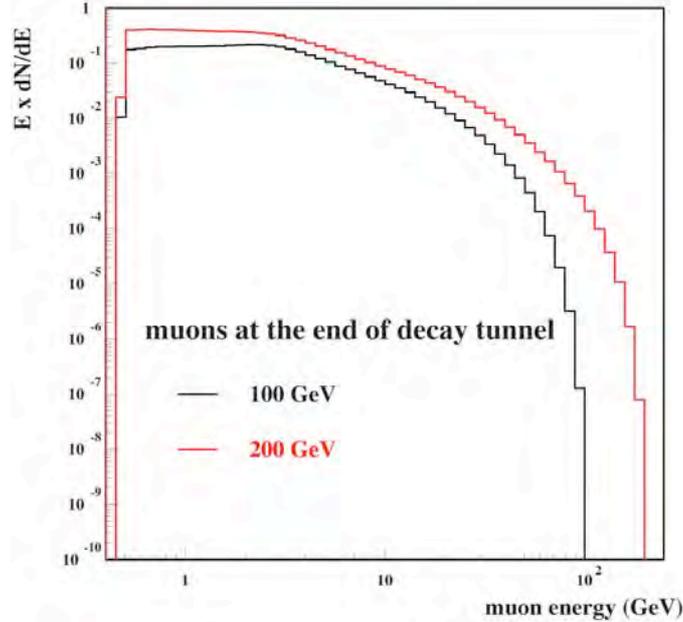

**Figure 2:** Muon lethargy spectra at the end of a ~ 100 m decay path, calculated for a 100 GeV (black curve) and 200 GeV (red curve) proton beam.

The neutrino event rates scale almost linearly with the beam energy, therefore the requirements of the experiment can be better quantified in terms of integrated beam power rather than in terms of pot. In the following tables and plots, the integrated intensity delivered to the CNGS facility, $4.5 \times 10^{19}$ pot/year at 100 GeV, has been assumed as a conservative reference.

A fast proton extraction from SPS is also a requirement for the LAr-TPC operation at surface in order to time tag the beam related events among the cosmic ray background.

The resulting neutrino CC event spectra peak at 2 GeV, with a broad energy range and a tail extending to higher energies (Figure 3). The similarity of the neutrino spectra at the near and far positions is excellent for $\nu_e$ and fairly good for $\nu_\mu$. The event rates at the Far and Near detectors with an instrumented LAr mass of 476 and 119 tons, respectively, are reported in Tables 1 and 3 (rows 1 to 3) for the negative and positive beam polarity, respectively. The reconstructed muon neutrino CC event rates for the two Near and Far iron mag-nets of 661 and 241 tons fiducial mass[1], respectively, are reported in Tables 2 and 4 (rows 1 to 4).

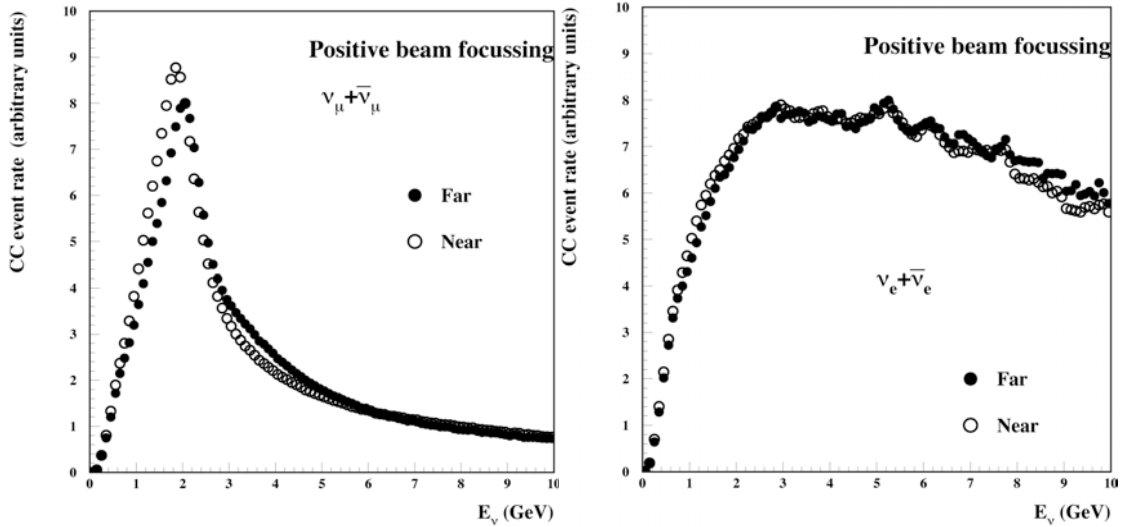

**Figure 3:** Muon (left) and electron (right) neutrino CC interaction spectra, at the Near and Far positions, arbitrary normalization.

Given the present phenomenological scenario, it is preferable to run first for e.g. 1+1 years in the antineutrino mode and then 1 year in the neutrino mode. The estimations in the following tables and the sensitivities in the related figures were therefore made for those running periods.

Finally, we note that, by assuming an intensity of $2 \times 10^{13}$ protons/spill, at the same level of CNGS, about 5 and 0.65 neutrino interactions per spill are expected in the Near and Far detectors, respectively, for positive focusing. Similarly 2.5 and 0.35 events per spill are expected in the two detectors for negative focusing. These rates are well within the capabilities of the two DAQ systems, for the two LAr-TPC and the two Spectrometers.

**Table 1.** Interaction rates in the LAr-TPC's at the Near (119 t) and Far locations (476 t) for 1 year of operation ($4.5 \times 10^{19}$ pot) with negative polarity beam. Oscillation events, in the electron-neutrino appearance mode, are also shown for some $\Delta m^2 - \sin^2(2\theta)$ parameters.

| | Near | | Far | |
|---|---|---|---|---|
| | Total | E<6 GeV | Total | E<6 GeV |
| $\nu_\mu$ CC | $2.49 \times 10^6$ | $7.26 \times 10^5$ | $3.44 \times 10^5$ | $9.66 \times 10^4$ |
| anti-$\nu_\mu$ CC | $1.88 \times 10^6$ | $1.31 \times 10^6$ | $2.42 \times 10^5$ | $1.69 \times 10^5$ |
| $\nu_e$ + anti-$\nu_e$ CC | $1.17 \times 10^5$ | $3.49 \times 10^4$ | $1.54 \times 10^4$ | $4.22 \times 10^3$ |
| $\Delta m^2 = 0.4$ eV², $\sin^2(2\theta) = 0.02$ | $3.41 \times 10^2$ | $3.32 \times 10^2$ | $8.17 \times 10^2$ | $7.84 \times 10^2$ |
| $\Delta m^2 = 2.0$ eV², $\sin^2(2\theta) = 0.002$ | $6.13 \times 10^2$ | $5.91 \times 10^2$ | $4.38 \times 10^2$ | $3.62 \times 10^2$ |
| $\Delta m^2 = 0.064$ eV², $\sin^2(2\theta) = 0.96$ | $4.32 \times 10^2$ | $4.21 \times 10^2$ | $1.46 \times 10^3$ | $1.42 \times 10^3$ |
| $\Delta m^2 = 4.42$ eV², $\sin^2(2\theta) = 0.0066$ | $6.01 \times 10^3$ | $5.67 \times 10^3$ | $1.89 \times 10^3$ | $9.89 \times 10^2$ |

---

[1] The magnet fiducial mass is defined vetoing events originating within 35 cm from the borders of the iron slabs in order to suppress contamination due to neutrino interactions outside the magnet.

**Table 2.** Rates of fully reconstructed events in the NESSiE spectrometers at the Near (241 t) and Far locations (661 t), for 1 year operation with negative polarity beam. The rows "NESSiE+LAr" correspond to events identified in both detectors. For some $\Delta m^2 - \sin^2(2\theta)$ «best fit» parameters the numbers of disappeared CC events are also shown (in parenthesis the percentage «oscillated/non-oscillated» is given).

| | Near | | Far | |
|---|---|---|---|---|
| | Total | E<6 GeV | Total | E<6 GeV |
| $\nu_\mu$ CC (NESSiE+LAr) | $1.19 \times 10^6$ | $2.29 \times 10^5$ | $1.42 \times 10^5$ | $2.09 \times 10^4$ |
| $\nu_\mu$ CC (NESSiE alone) | $2.32 \times 10^6$ | $1.15 \times 10^6$ | $1.86 \times 10^5$ | $9.41 \times 10^4$ |
| anti-$\nu_\mu$ CC (NESSiE+LAr) | $6.71 \times 10^5$ | $3.68 \times 10^5$ | $6.78 \times 10^4$ | $3.3 \times 10^4$ |
| anti-$\nu_\mu$ CC (NESSiE alone) | $1.52 \times 10^6$ | $1.13 \times 10^6$ | $1.19 \times 10^5$ | $8.98 \times 10^4$ |
| (3+1) $\Delta m^2 = 0.9$ eV$^2$, $\sin^2(2\theta)=0.083$ | $1.48 \times 10^3$ (0.1%) | $1.45 \times 10^3$ (0.13%) | $2.50 \times 10^3$ (2.1%) | $2.43 \times 10^3$ (2.7%) |
| (3+1) $\Delta m^2 = 1.6$ eV$^2$, $\sin^2(2\theta)=0.034$ | $1.87 \times 10^3$ (0.12%) | $1.84 \times 10^3$ (0.16%) | $1.98 \times 10^3$ (1.7%) | $1.70 \times 10^3$ (2.11%) |
| (3+2) $\Delta m_{41}^2 = 0.47$eV$^2$, $\Delta m_{51}^2 = 0.87$eV$^2$ | $1.96 \times 10^3$ (0.13%) | $1.92 \times 10^3$ (0.17%) | $3.50 \times 10^3$ (2.94%) | $3.42 \times 10^3$ (3.8%) |
| (3+2) $\Delta m_{41}^2 = 1.00$ eV$^2$, $\Delta m_{51}^2 = 1.60$ eV$^2$ | $3.68 \times 10^3$ (0.24%) | $3.6 \times 10^3$ (0.32%) | $5.12 \times 10^3$ (4.3%) | $4.96 \times 10^3$ (5.5%) |

**Table 3.** Interaction rates in the LAr-TPC's at the Near (300 m, 119 t) and Far locations (1600m, 476 t) for 1 year of operation (4.5 $10^{19}$ pot) with positive polarity beam. Oscillation events, in the electron-neutrino appearance mode, are also shown for some $\Delta m^2 - \sin^2(2\theta)$ parameters.

| | NEAR | | FAR | |
|---|---|---|---|---|
| | Total | E<6 GeV | Total | E<6 GeV |
| $\nu_\mu$ CC | $7.92 \times 10^6$ | $5.05 \times 10^6$ | $1.01 \times 10^6$ | $6.44 \times 10^5$ |
| anti-$\nu_\mu$ CC | $5.29 \times 10^5$ | $1.98 \times 10^5$ | $7.34 \times 10^4$ | $2.63 \times 10^4$ |
| $\nu_e$ + anti-$\nu_e$ CC | $1.60 \times 10^5$ | $5.45 \times 10^4$ | $2.07 \times 10^4$ | $6.44 \times 10^3$ |
| $\Delta m^2 = 0.4$ eV$^2$, $\sin^2(2\theta)=0.02$ | $1.09 \times 10^3$ | $1.08 \times 10^3$ | $2.47 \times 10^3$ | $2.42 \times 10^3$ |
| $\Delta m^2 = 2.0$eV$^2$, $\sin^2(2\theta)=0.002$ | $1.93 \times 10^3$ | $1.90 \times 10^3$ | $1.02 \times 10^3$ | $9.14 \times 10^2$ |
| $\Delta m^2 = 0.064$eV$^2$, $\sin^2(2\theta)=0.96$ | $1.37 \times 10^3$ | $1.36 \times 10^3$ | $4.47 \times 10^3$ | $4.42 \times 10^3$ |
| $\Delta m^2 = 4.42$eV$^2$, $\sin^2(2\theta)=0.0066$ | $1.75 \times 10^4$ | $1.70 \times 10^4$ | $3.59 \times 10^3$ | $2.40 \times 10^3$ |

**Table 4.** Rates of fully reconstructed events in the NESSiE spectrometers at the Near (241 t) and Far locations (661 t), for 1 year operation with positive polarity beam. The rows "NESSiE+LAr" correspond to events identified in both detectors. For some $\Delta m^2 - \sin^2(2\theta)$ «best fit» parameters the numbers of disappeared CC events are also shown (in parenthesis the percentage «oscillated/non-oscillated» is given).

| | NEAR | | FAR | |
|---|---|---|---|---|
| | Total | E<6 GeV | Total | E<6 GeV |
| $\nu_\mu$ CC (NESSiE+LAr) | 2.78 x10$^6$ | 1.22 x10$^6$ | 2.86 x10$^5$ | 1.06 x10$^5$ |
| $\nu_\mu$ CC (NESSiE alone) | 5.35 x10$^6$ | 3.59 x10$^6$ | 4.15 x10$^5$ | 2.84 x10$^5$ |
| anti-$\nu_\mu$ CC (NESSiE+LAr) | 1.89 x10$^5$ | 5.62 x10$^4$ | 3.15 x10$^4$ | 6.9 x10$^3$ |
| anti-$\nu_\mu$ CC (NESSiE alone) | 3.99 x10$^5$ | 2.98 x10$^5$ | 4.33 x10$^4$ | 2.25 x10$^4$ |
| (3+1) $\Delta m^2$ =0.9 eV$^2$, $\sin^2(2\theta)$=0.083 | 3.76 x10$^3$ (0.07%) | 3.18 x10$^3$ (0.1%) | 6.44 x10$^3$ (1.55%) | 6.25 x10$^3$ (2.2%) |
| (3+1) $\Delta m^2$ =1.6 eV$^2$, $\sin^2(2\theta)$=0.034 | 4.76 x10$^3$ (0.13%) | 4.69 x10$^3$ (0.16%) | 5.26 x10$^3$ (1.27%) | 5.03 x10$^3$ (1.77%) |
| (3+2) $\Delta m_{41}^2$ =0.47eV$^2$, $\Delta m_{51}^2$ =0.87eV$^2$ | 4.99 x10$^3$ (0.1%) | 4.92 x10$^3$ (0.14%) | 9.03 x10$^3$ (2.17%) | 8.77 x10$^3$ (3.08%) |
| (3+2) $\Delta m_{41}^2$ =1.00 eV$^2$, $\Delta m_{51}^2$ =1.60 eV$^2$ | 9.37 x10$^3$ (0.17%) | 9.18 x10$^3$ (0.25%) | 1.33 x10$^4$ (3.21%) | 1.29 x10$^4$ (4.53%) |

### 1.2  Expected sensitivities to neutrino oscillations.

A complete discussion of $\nu_\mu \rightarrow \nu_e$ oscillation studies both in appearance and disappearance modes has been presented in the SPSC-P345 document [2], that includes the genuine event selection and background rejection in the LAr-TPC's,. In particular, due to the excellent $\pi^0$–to-electron separation, a $\pi^0$ rejection at $10^3$ level is obtained when requiring at least 90 % electron recognition efficiency.

The effects of the high-energy event tail in the event selection has been studied on fully simulated events. Muon neutrino CC interactions in LAr-TPC can fake electron neutrino events when the muon is not correctly recognized and at least one $\pi^0$ is misidentified as an electron. To this aim, it has been conservatively assumed that a μ can be misidentified as a π, if it escapes the detector leaving a track shorter than 250 cm. For the $\pi^0$ identification the same procedure and the same fiducial cuts as for NC events apply. The resulting background is very small, of the same order of the residual NC background. The visible energy spectra are shown in Figure 4 for one set of the oscillation parameters. In Tables 1 and 3 (rows 4 to 7) the expected event rates for the $\nu_e$ appearance in the instrumented LAr detector are reported for 4 different parameter sets of sterile neutrino model.

In addition to the $\nu_\mu \rightarrow \nu_e$ oscillation studies mentioned above, $\nu_\mu$ oscillation in disappearance mode was discussed at length in [3], by using large mass spectrometers with high capabilities in charge identification and muon momentum measurement. It is important to note that all sterile models predict large $\nu_\mu$ disappearance effects together with $\nu_e$ appearance/disappearance. To fully constrain the oscillation searches, the $\nu_\mu$ disappearance studies have to be addressed. Much higher disappearance probabilities are expected than in appearance mode, where relative amplitudes as large as 10% are possible. The spectrometers will be able to correctly identify about 40% of all the CC events produced in, and escaped from, the LAr-TPC's, both in the near and far sites. That will increase the fraction of CC events with charge identification and momentum measurement, especially at high energies. Therefore complete measurement of the CC event spectra will be possible, along with the NC/CC event ratio (in synergy with the LAr-TPC), and the relative background systematics.

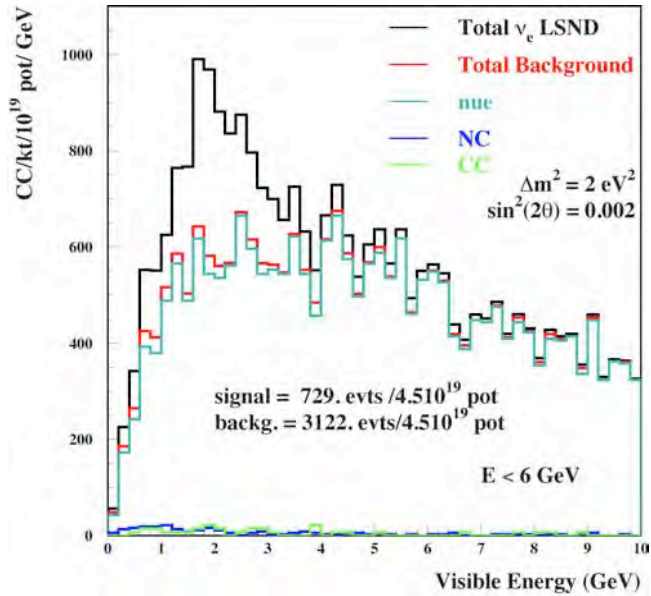

**Figure 4:** Spectra of the visible energy in the Far detector for $\nu_e$ -CC like events, for a set of oscillation parameters. The total spectrum with oscillation (black) is compared to the background (red). The background components are the intrinsic $\nu_e$ + anti-$\nu_e$ events (green), the misidentified NC events (blue) and the misidentified CC events (yellow). The event rates are for 4.5 $10^{19}$ pot at 100 GeV in the T600 detector with a 90% efficiency (see text for details).

The large mass of the magnets will also allow an internal check of the NC/CC ratio in an extended energy range, and an independent measure of the CC oscillated events. As an example, the expected spectra for the measured CC muon events for the non-oscillation and the oscillation hypothesis, are shown in Figure 5, for the "NESSiE alone" detection.

In Tables 2 and 4 (rows 5 to 8) the expected disappeared CC events for some "best fits" of sterile neutrino models in the spectrometer fiducial masses are reported for 4 different parameter sets of sterile neutrino model, Near and Far locations, negative and positive polarity beams, respectively.

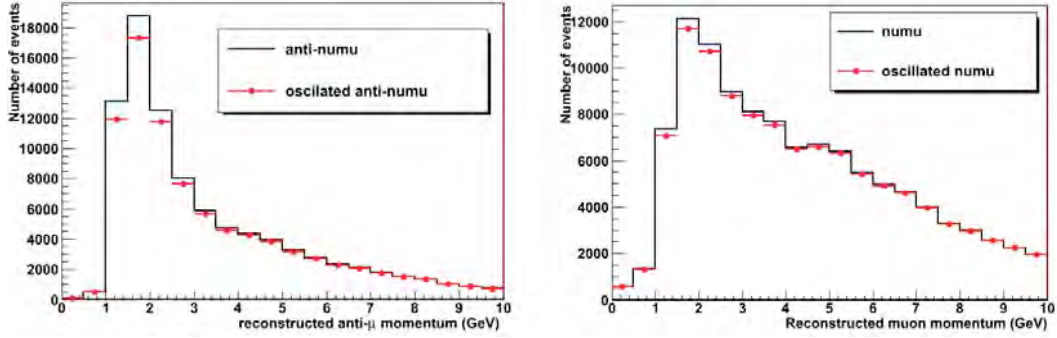

**Figure 5:** Spectra of the measured muon momenta in the Far detector for anti-$\nu_\mu$ CC events, for the last set of oscillation parameters of Tables 2. The total spectrum with oscillation (red) is compared to the non-oscillated distribution. The event rates are for 4.5x10^19 pot at 100 GeV, negative polarity reconstructed in the fiducial volumes of the Spectrometer alone. It is noticeable that with 1 year CNGS-like running in inverse polarity, both antineutrino and neutrino exploitations can be made.

### 1.2.1 $\nu_\mu \rightarrow \nu_e$ appearance signal

A sensitivity of $\sin^2 2\theta_{new} > 3 \times 10^{-4}$ (for $|\Delta m^2_{new}| > 1.5$ eV$^2$) and $|\Delta m^2_{new}| > 0.01$ eV$^2$ (for $\sin^2 2\theta_{new} = 1$) at 90 % C.L. is expected for the $\nu_\mu \rightarrow \nu_e$ transition with one year exposure (4.5x10^19 pot) at the CERN-SPS $\nu_\mu$ beam (Figure 6-left). The parameter space region allowed by the LSND experiment is fully covered, except for the highest $\Delta m^2$ region. The sensitivity has been computed according to the above described particle identification efficiency and assuming a 3% systematic uncertainty in the prediction of "Far" to "Near" $\nu_e$ ratio. A further control of the overall systematics will be provided by the LAr and spectrometer combined measurement of CC spectra in the Near site and over the full energy range.

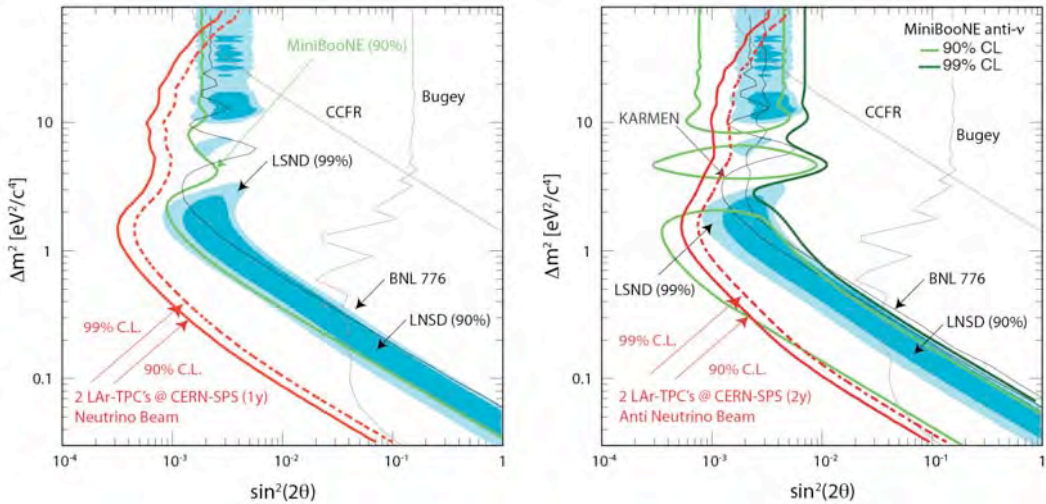

**Figure 6:** Expected sensitivity for the proposed experiment exposed at the CERN-SPS neutrino beam (left) and anti-neutrino (right) for 4.5 10^19 pot (1 year) and 9.0 10^19 pot (2 years) respectively. The LSND allowed region is fully explored in both cases

In anti-neutrino focusing, twice as much exposure (0.9x10^20 pot) allows to cover both the LSND region and the new MiniBooNE results (Figure 6-right) [5]. Both favoured MiniBooNE parameter sets, corresponding to two different energy regions in the MiniBooNE antineutrino analysis, fall well within the reach of this proposal.

### 1.2.2 $\nu_e$ and $\nu_\mu$ disappearance signals

In Figure 7 the sensitivity for $\nu_e$ disappearance search in the $\sin^2(2\theta_{new})$ - $\Delta m^2_{new}$ plane is shown for the presently proposed experiment with an integrated intensity of $4.5\times10^{19}$ pot, corresponding to two years of data taking at the presently available beam intensity. The oscillation parameter region related to the "anomalies" from the combination of the published reactor neutrino experiments, Gallex and Sage calibration sources experiments [6] is completely explored.

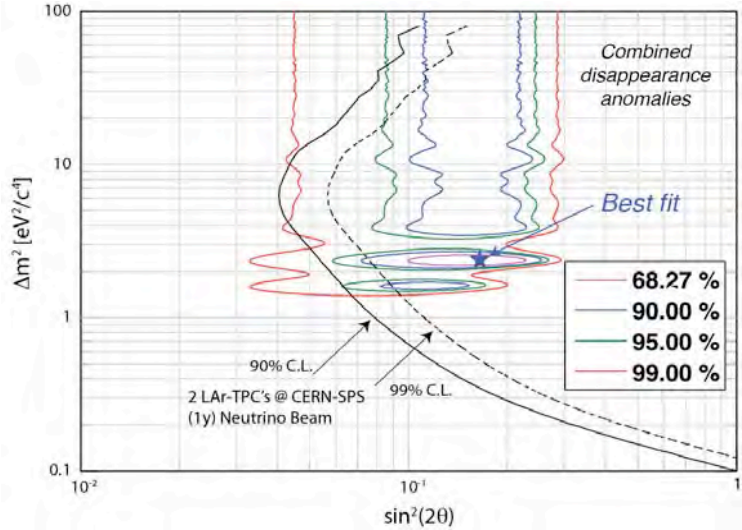

**Figure 7:** Oscillation sensitivity in the $\sin^2(2\theta_{new})$ vs. $\Delta m^2_{new}$ distribution for the CERN-SPS neutrino beam (1 year). A 1% overall and 3% bin-to-bin systematic uncertainty on the energy spectrum is included. Combined disappearance "anomalies" from the published reactor neutrino, Gallex and Sage calibration sources experiments are also shown.

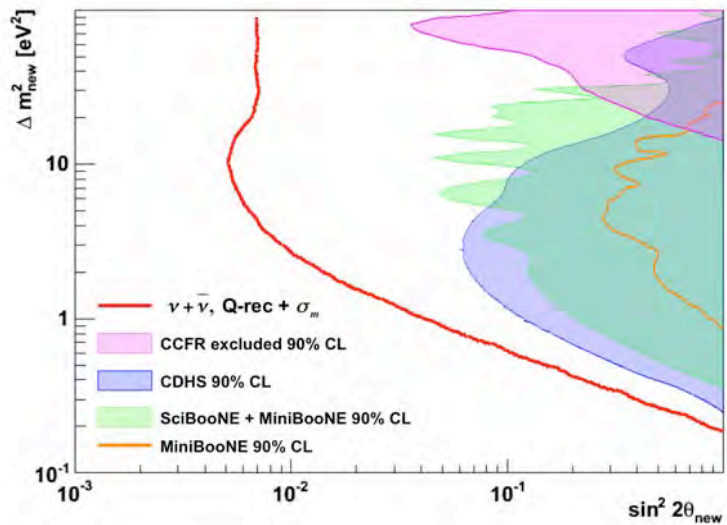

**Figure 8:** The sensitivity plot (at 90% C.L.) for the negative-focusing option considering three years of the CERN-SPS beam (2 years in anti-neutrino and 1 year in neutrino mode) from CC events fully reconstructed in NESSiE+LAr. Red line: $\nu_\mu$ exclusion limit. The three filled areas correspond to the present exclusion limits on the $\nu_\mu$ from CCFR, CDHS and SciBooNE+MiniBooNE experiments (at 90% C.L.). Orange line: recent exclusion limits on $\nu_\mu$ from MiniBooNE alone measurement.

Moreover the $\nu_\mu$ disappearance signal is well studied by the spectrometers, with a very large statistics and disentangling of $\nu_\mu$ and $\bar{\nu}_\mu$ interplay [7]. As an example, Figure 8 shows the sensitivity plot (at 90% C.L.) for two years negative-focusing plus one year positive-focusing. A large extension of the present limits for $\nu_\mu$ by CDHS and the recent SciBooNE+MiniBooNE will be achievable in $\sin^2(2\theta) - \Delta m^2$.

Both $\nu_e$ and $\nu_\mu$ disappearance modes will be used to add "conclusive" information on the sterile mixing angles as shown in the following Table 5.

**Table 5.** The measurements on the mixing angle as provided by different experiments.

| Oscillation type | Neutrinos | Experiments |
|---|---|---|
| $\theta_{12}$ | $\nu_e$(solar, reactors) | SNO, SK, Borexino, Kamland |
| $\theta_{23}$ | $\nu_\mu$ (atmospheric, accelerators) | SK, Minos, T2K |
| $\theta_{13}$ | $\nu_e$(reactors) | Daya Bay, Double Chooz |
| $\theta_{14}$ | $\nu_e$(reactors, radioactive sources) | SBL Reactors, Gallex, Sage. **This Proposal** |
| $\theta_{24}$ | $\nu_\mu$ (accelerators) | CDHS, Miniboone. **This Proposal** |

## 2   The LAr-TPC detectors.

An essential feature of the experiment is the close identity of the two LAr detectors in order to ensure the equality of the observed neutrino event spectra in the absence of oscillations.

The T600 detector will be moved from the LNGS laboratory into the Far position of the CERN neutrino beam line. This will be done with minimal changes, using most of the existing equipment. In view of its new location, some components will be renewed: cold vessels, thermal shields and external insulation, phototubes.

The T600 internal detector will be extracted fully assembled from the cryostat at LNGS into a clean room container protecting and supporting the chambers for the transport to CERN. Several components of the detector and of the associated cryogenic equipment must be disassembled, like for instance the DAQ electronics, the cryogenic storage, the auxiliary plants and the $LN_2$ re-liquefaction system. The present light collecting system will be replaced with higher performance phototubes and installed within the cryostats.  The whole detector will be assembled at CERN.

A new LAr-TPC detector, but of a smaller size (T150), will be implemented in the near location at a shorter distance from the beam target. The T150 will be an exact "clone" of one T600 module, with a reduction of a factor 2 (about 10 m) in length. The additional LAr mass will be 200 t (119 t active mass). The same type of wire chambers, mechanics, wire planes and the 1.5 m maximum drift distance will be preserved. The cold vessel, the insulation vessel and the cryogenics equipment will also be consistent with the solutions adopted for T600 at CERN.

### 2.1   *General description of the T600 detector*

The ICARUS T600 detector [8], presently fully operational in the LNGS Hall B, consists of a large cryostat split into two identical, adjacent modules with internal dimensions 3.6 x 3.9 x 19.6 $m^3$ filled with about 760 tons of ultra-pure liquid Argon (Figure 9). Such units are operated together as a unique detector. However in view of the CERN programme, the possibility of operating the two modules separately at different distances is under consideration.

A uniform electric field ($E_D = 500$ V/cm) is applied:  each module houses two TPCs separated by a common cathode.  Each TPC is made of three parallel planes of wires, 3 mm apart, facing the drift path (1.5 m). Globally, 53248 wires with length up to 9 m are installed in the detector. By appropriate voltage biasing, the first two signal sensing planes (Induction-1 and Induction-2) provide differential signals in a non-destructive way, whereas the ionization charge is finally collected by the last Collection plane.

On each plane, wires are oriented at 0°, ±60° angles with respect to the horizontal direction. Therefore a three-dimensional image of the ionizing event is reconstructed combining the wire coordinate on each plane at a given drift time. A remarkable resolution of about 1 mm³ is uniformly achieved over the whole detector active volume (~340 m³ corresponding to 476 t).

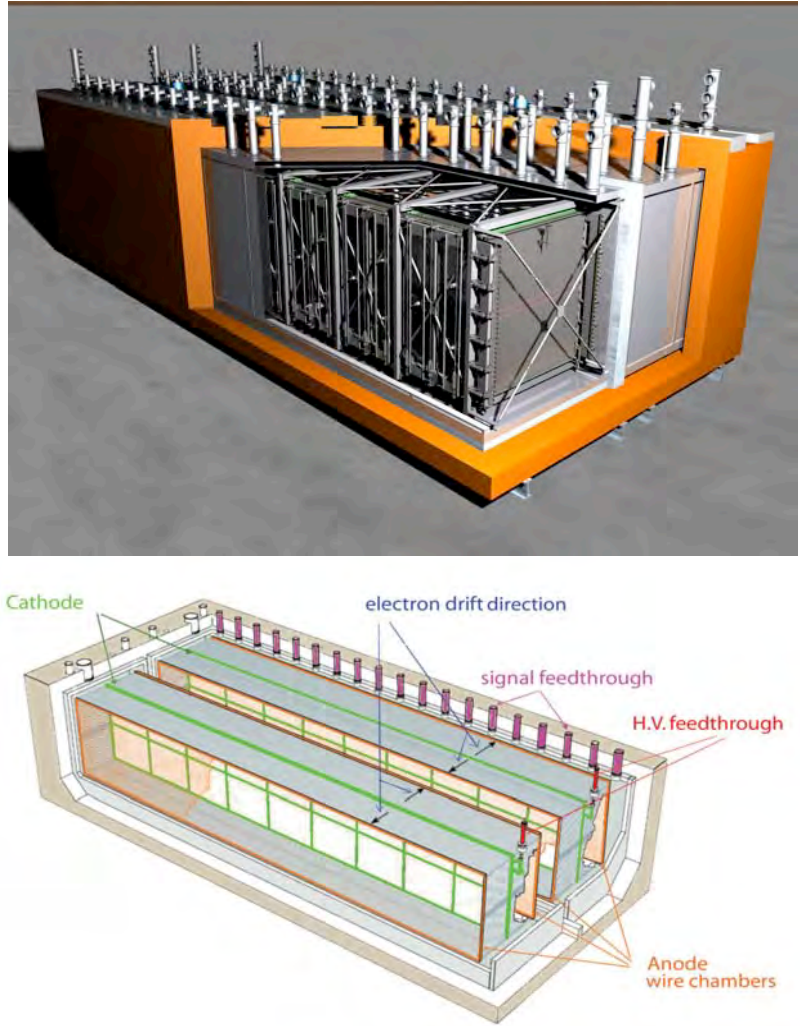

**Figure 9:** The ICARUS T600 detector schematics, showing both half-modules and the common insulation surrounding the detector; inner structures and feed-throughs are also shown. On the bottom a view of the detectors is presented with the wire chambers and the high voltage system (race-tracks and cathodes).

The measurement of the absolute time of the ionizing event, combined with the electron drift velocity information ($v_D \sim 1.6$ mm/$\mu s$ at $E_D = 500$ V/cm), provides the absolute position of the track along the drift coordinate. The absolute time of the ionizing event is given by the prompt detection of the scintillation light produced in LAr by ionizing particles. An array of several Photo Multiplier Tubes (PMTs) installed behind each of the wire chambers and coated with wavelength shifter allows the detection of VUV scintillation light ($\lambda = 128$ nm), at a LAr cryogenic temperature.

All the detector components, including PMTs, cables, monitors and probes are held by a self-supporting low carbon stainless steel structure. The structure is coupled to the floor of the aluminium vessels by 10 supports. One of them is fixed, while the others are sliding to allow for the relative movement of the detector structure and the aluminium vessels during the cooling phase due to thermal shrinking.

Stainless-steel chimneys are aligned in two rows on the aluminium ceiling of each module (20 per row) and terminated with special vacuum-tight feed-through flanges (INFN patent RM2006A000406). They insure the passage of the wire signal cables (576 per flange) and of other detector subsystem (PMT's) and the control of the instrumentation. Each flange is connected to a single electronic rack where the front-end electronic, the digitisers and the memory buffers are hosted for the readout of 572 channels.

The electronics was designed to allow the continuous read-out, digitization and independent waveform recording of the signals from each wire of the TPC. The read-out architecture consists of a front-end low noise charge sensitive pre-amplifier, allowing a signal-to-noise better than 10:1. Signals coming from each wire are independently digitized every 400 ns with the help of a 10 bit FADC. This scheme is implemented on a single VME-like analogue board hosting 32 channels amplifiers, multiplexers, ADC's and a 20-bit, 40 MHz serial link. Data are sent to a digital board for buffering.

A thermal insulation vessel, made with evacuated honeycomb panels and assembled to realize a tight containment, surrounds the two modules. In order to intercept the heat load and maintain the cryostat's bulk temperature of 89 K within 1 K, a thermal shield with boiling Nitrogen circulating inside, is located between the insulation and the aluminium containers.

Nitrogen, used to cool the whole T600, is stored in two 30 $m^3$ $LN_2$ tanks. Its temperature is fixed by the equilibrium pressure in the tanks (~ 2.1 bar, corresponding to about 84 K), which is kept stable in steady state. A dedicated re-liquefaction system (twelve cryo-coolers, 48 kW total cold power) is installed ensuring a safe operation in closed-loop.

As a fundamental requirement of a LAr-TPC, free electrons produced by ionizing particles must travel "unperturbed" from the point of production to the wire planes. Electronegative impurities (mainly $O_2$, $H_2O$ and $CO_2$) must be kept at a remarkably low concentration level (less than 0.1 ppb, corresponding to a free electron lifetime of several milliseconds). Each module is equipped with two gas-argon and one liquid-argon recirculation/purification systems. Argon gas is continuously drawn from the cryostat ceiling and — once re-condensed — drops into Oxysorb ™ filters and returns as LAr. An immersed, cryogenic pump purifies the LAr through standard Hydrosorb/Oxysorb™ filters, before re-injecting into the cryostats, insuring the recirculation of the full volume in about 6 days.

During the filling phase, cryostats have been carefully evacuated ensuring an appropriate out-gassing of the internal walls of the cryostat and of all the detector materials. The electron lifetime obtained in the T600 module exceeds 6 ms as shown in Figure 10, guaranteeing a signal attenuation of at most 13% for the maximum electron drift time of 1ms.

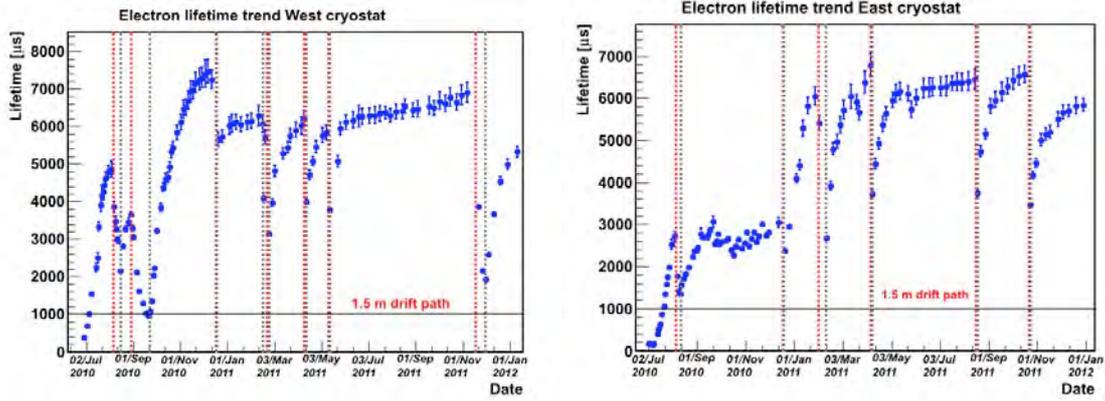

**Figure 10.** Time evolution of the free electron lifetime in the West (left) and East (right) cryostats at LNGS. continuously measured by means of the charge attenuation along crossing muon tracks. The electron lifetime is constantly above 6 ms in both cryostats, when the liquid recirculation system is active at a rate of 2 m$^3$/hour, This corresponds to a maximum free electron yield attenuation of 14 %, at the maximum drift distance of 1.5 m (last reference in [8]).

### 2.2    T600 transport from LNGS to CERN

A feasibility study and cost estimate for the possible transport of ICARUS T600 from LNGS Hall B to CERN has been carried out in close collaboration with the Transport Group of CERN, avoiding any major heavy works in LNGS underground Laboratory.

The decommissioning procedures of the T600 detector will start by turning off high voltage and all the readout electronics. LAr recirculation will be stopped, while the gas recirculation will continue to run at decreasing rates in order to maintain the LAr pressure. The nitrogen cooling system will continue to run until the argon vessels are completely empty. Emptying of the argon vessels will proceed by transferring the LAr to trucks, with the same procedure used in 2001, at the end of the technical run in Pavia. According to the previous experience about 85% of the LAr will be recovered. At the end of the emptying process, venting valves will be opened and vacuum in the insulation panels will be broken to speed up the evaporation of the residual argon and warming up. This process is expected to take about two months, during which disassembly of electronics/DAQ, and part of the nitrogen and argon circuits, can start.

Access to the internal detector implies the removal of (1) the two LN$_2$ storage vessels; (2) the top external supporting structure; (3) the interfering part of N$_2$ and Ar pipes; (4) the gas recirculation systems on the front side; (5) LAr and LN$_2$ transfer lines on the front side.  Moreover the external containment cage and the external insulation of the front side will be disassembled to cut the front containment wall, to remove the liquid nitrogen shield on the front side and finally to cut the welded sleeves to allow the opening of the front door of the aluminium vessel.

In order to avoid exposure of the internal detector to the gallery atmosphere, before the removal of the front door of the cold vessels, an adequate "box" will be positioned in front of the opening to be connected by a flexible joint to the vessel immediately after the door removal. Highly filtered air (compatible with 10000 class clean room specifications) will start to be flushed from the rear side of the cold vessel to the front of the external container. The volume resulting from the connection of external container to the cold vessel will be an equivalent class clean room.

In order to have access to the internal detector supports part of the race-tracks need to be dismounted (Figure 11). One third of the race-tracks in the front and bottom sides of the detector will be removed.

There are 10 bolted and pinned supports between the internal detector structure and the cryostat floor: one fixed in the middle of the structure and the others sliding. To extract the internal detector all the supports need to be un-bolted (Figure 12). In parallel with these operations a number of other actions will be performed on top of the vessel: (1) removal of all remaining cryogenic equipment (including pumps, purifiers and pipes), top thermal insulation, cooling screen straight chimneys and the upper flanges; (2) disconnection from the feed-throughs of internal cables (packaged will be placed on top of the internal detector structure) and auxiliary internal instrumentations; (3) dismounting of the cross chimneys. All these operations will be carried with the precautions required in order to prevent contamination of the internal detector components.

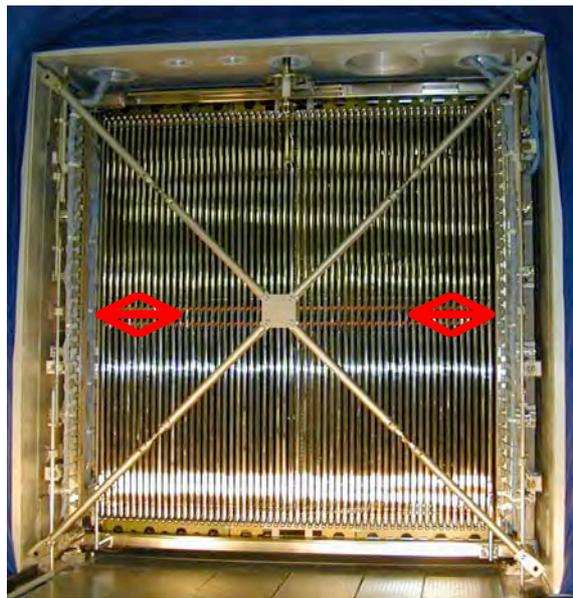

**Figure 11**: T600 internal detector as seen from the front opening of the cold vessel; the race tracks in the two regions indicated by the red arrows will be dismounted in order to access the detector supports.

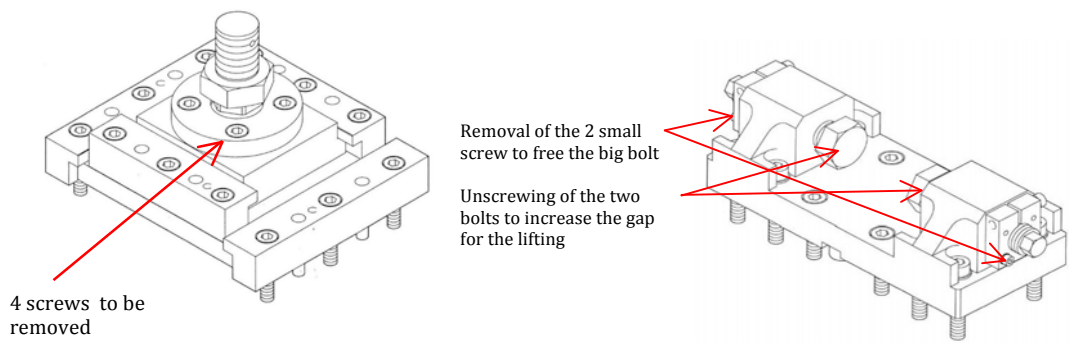

**Figure 12**: Disconnection of the fixed and sliding supports.

Once the supports have been unbolted, the whole internal detector will be lifted and moved outside the vessel with the same procedure already used for the insertion: (1) rails will be placed on the cryostat floor and on the outside "box" (Figure 13); (2) an existing set of 10 hydraulic lifters, with wheels, will be placed below the internal detector structure close to the supports; (3) two electrical engines will be attached at the front of the internal detector structure. The whole structure will be then lifted by about 50 mm and pulled outside the vessel into the transport "box" made of alloy sheets (5 mm thickness) mounted on a "transport frame" to ensure mechanical protection and clean environment for the TPCs during transport and storage in the parking positions.

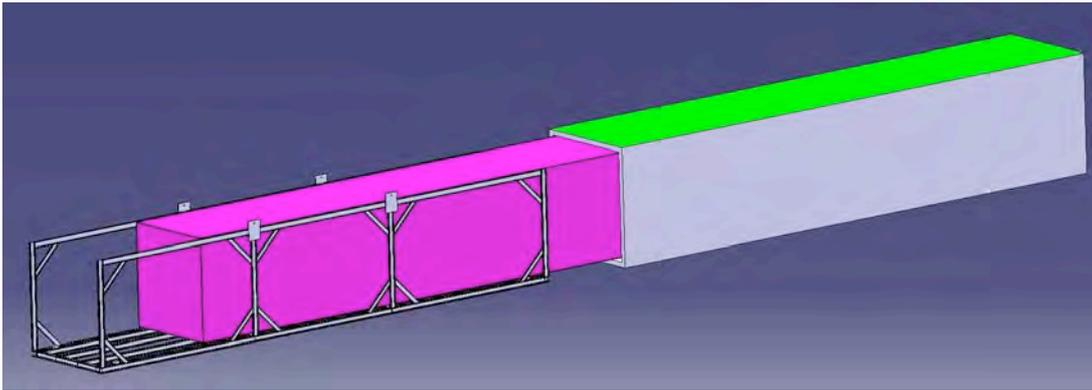

**Figure 13**: Schematic drawing illustrating the transfer of the internal detector from the aluminium vessel into the "box" for clean storage and transportation.

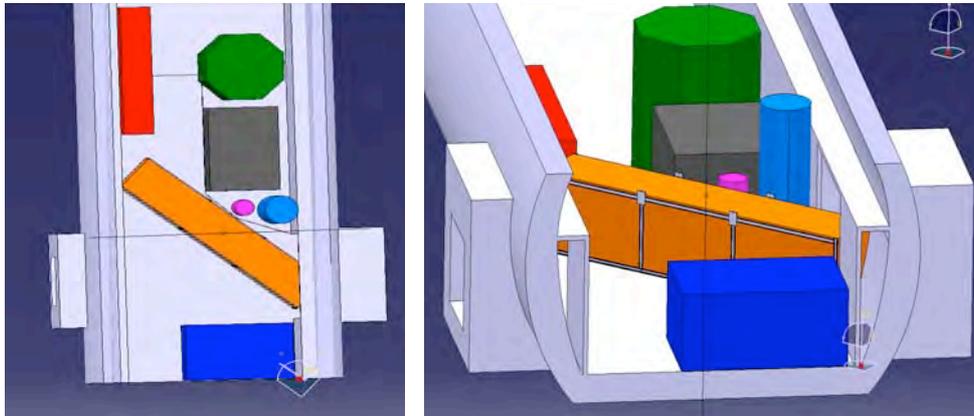

**Figure 14**: 3D simulation of the internal detector extraction from Hall B (left: top view, right: front view).

Once the internal detector is placed inside the transport frame, it will be disconnected from the vessel and immediately sealed on the rear side to preserve cleanliness. The internal detector, inside the transport box will be moved to the entrance of Hall B. Since the overall weight of the internal detector + transportation box is well below the maximum load of the Hall B crane (40 ton), this operation can be performed using the existing crane. A 3D simulation shows that the extraction of the internal detector (inside the box) can be performed without the need of significant dismount of the equipment and infrastructures presently installed in LNGS (Figure 14). When the internal detector will be at the entrance of Hall B, oriented along the TIR gallery, the two trolleys for transport will be connected to the "box" (Figure 15). The convoy will exit from the underground laboratory from the entry side and it will start its route to CERN.

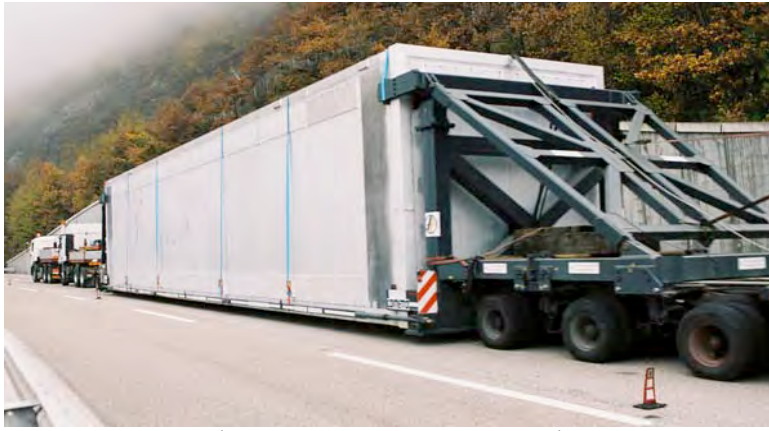

**Figure 15**: Picture taken during the transportation of a T600 half-module to LNGS. One of the two trolleys, attached to the aluminium vessels rear side, is well visible.

Upon arrival, the transport box will be disconnected from the trolleys and the internal detector will be transferred in the intermediate working area. The transport box will then be available for the second module. A second transportation box will be needed if the first one will be used as a temporary storage. The availability of two transport boxes, serving also as temporary storage, will make the assembly organization and scheduling much more flexible.

All the remaining components of the T600 plant will be disassembled and appropriately packaged and transported: (1) all the readout electronics, the electric cabinets, HV and LV power supplies, DAQ system, slow control systems; (2) the T600 $LN_2$, LAr and GAr systems (nitrogen shields, transfer lines, pumps, valves, auxiliary reservoirs, gas and liquid recirculation units, purifiers); (3) the power supplies for the cryogenic plant, the automatic control system cabinet; (4) the Stirling re-liquefaction system (3 skids with 4 units each), including auxiliary equipment; (5) $LN_2$ and LAr storage dewars (2 x 30000 litres each).

The external support structure (including seismic feet) will also be moved for a possible re-use at CERN. Transport will be performed with 2 exceptional convoys for internal detector containers; 2 open convoys for tanks; 3 closed convoys for pumping/compressor groups. The bulk material will require additional 15-20 cargo convoys.

### 2.3    *The new cold vessels.*

The LAr will be contained in three mechanically independent vessels, two of about 270 m$^3$ each for the T600 detector and one of 165 m$^3$ for the T150. According to the present experience, to outgas efficiently the internal surfaces and obtain an appropriate LAr purity, the cold vessels must be evacuated to less than $10^{-3}$ mbar. Therefore the vessels need to stand the vacuum and to be tight to better than $10^{-5}$ mbar l sec$^{-1}$. Moreover they will be designed for a maximum operating internal overpressure of 1 bar (0.45 bar relief valve settings + 0.55 bar hydrostatic pressure).

The new T600 vessels will be of parallelepiped shape with internal dimensions 3.6 (w) x 3.9 (h) x 19.3 (l) m$^3$ to match the existing internal detector.

The T150 vessel will have the same cross-section (internal 3.6 (w) x 3.9 (h) m²; external 3.9 (w) x 4.2 (h) m²) and an internal length of 11.8 m (12.1 m external).

They will be realized by welding together extruded aluminium profiles (Figure 16 and 17) with a significant simplification with respect to the aluminium honeycomb used in the present detector, at the cost of a slight increase in the cryostat weight (30 t each). The use of aluminium LAr vessels is also particularly attractive in view of the very good shielding offered against external electronic noises and the large thermal conductivity that improves the temperature uniformity inside the LAr. Nowadays there are industries very qualified and experienced in the construction of aluminium vacuum tight containers, as demonstrated also by the LHC experience.

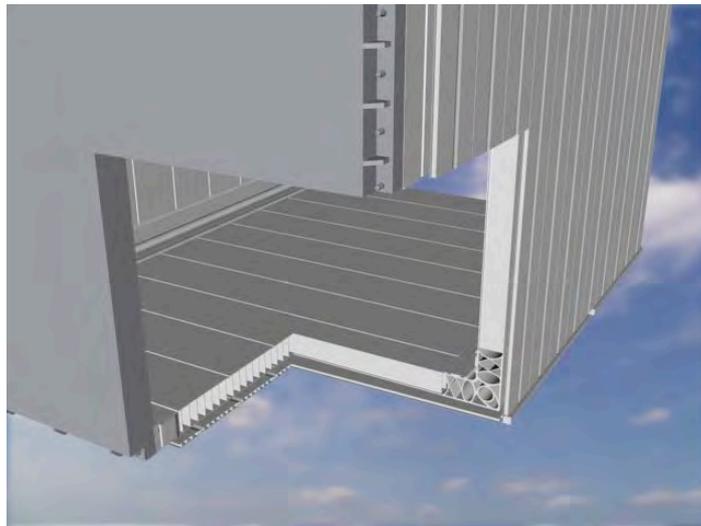

**Figure 16**: Rendered view of a cross-section of the new aluminium vessels. The external cooling shield is also visible.

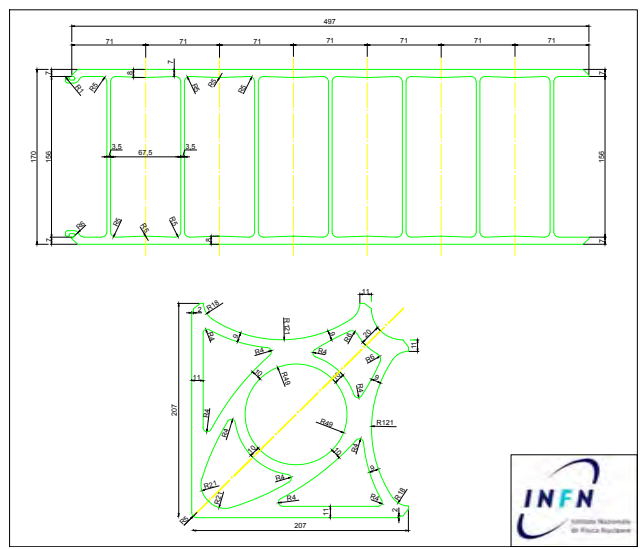

**Figure 17**: Technical drawing of some aluminium extrusions for the construction of the new LAr vessels.

The external size and weight of the new vessels allow for their transport from the manufacturer to CERN. The amount of construction work to be done in CERN will be therefore minimized (vessels opening for the internal detector installation and subsequent closure) with significant cost and time saving. This design will take advantage of the previous studies developed for the T1800 INFN program, almost at the level of executive design, especially for cryogenics and internal detector mechanics. This new design will apply to both the new cold vessels and the internal detector of the T150 module.

As in the present T600 vessels, the double-layered walls can be evacuated independently from the main LAr volume, leading to efficient leak detection and repair, during the construction phase. In addition this solution adds an extra safety margin when the vessels are under vacuum before cooling and filling with LAr. If, also in this phase, the vessel walls are evacuated, the walls stiffness is increased and a barrier is created against possible, residual, external leak.

The angular profiles of the new vessels have a set of four rails. Special aluminium blocks will be inserted in the vessel floor in correspondence to the existing internal detector supports. These elements will be machined, after completion of the vessel assembly, to attain the alignment required by the existing detector mechanics.

On the top of the T600 vessels, there are 36 openings with internal diameter 200 mm, each one serving the readout cables from 576 of the wires at ± 60°, equally spaced at about 1 m from each other and organized in two rows located near the edges of the top wall (just above the wire chambers). Four additional 250 mm diameter openings near the corners of the top wall host the cables for the horizontal wires and for the last portion of wires at ± 60° (those with degrading length). On the top are also the LAr inlet, for GAr recirculation and other services (vacuum, safety disks, instrumentation, manholes). The output port to the LAr recirculation system is approximately at the centre of the backward end cap. Finally, an aperture in the backward bottom edge is the port for the vessel emptying.

The T150 vessel will have 16 x 200 mm diameter plus 4 x 250 mm diameter ports for the TPC signal cables. One port only is foreseen for a GAr recirculation unit and one for the safety disk. All the remaining openings will be identical to the ones of a T600 module.

Due to the severe requirements on vacuum tightness, we decided to use for the argon circuit ConFlat flanges (Figure 18). On all the output ports of the main LAr containers, excluding the manholes, special vacuum tight bi-metal (aluminum to stainless steel) joints are installed with the inner aluminium part welded to the container body and the outer stainless steel part machined as a standard ConFlat flange. On these flanges, ConFlat joints, 800 mm long, are installed, to move the ports outside the insulation layer. Signal feedthrough flanges are therefore installed outside the insulation at room temperature. The container is transported sealed at the level of the bi-metal joints, without the additional connections. For the manholes sealing is made with an aluminium flange with helicoflex gaskets.

For the new T600 vessels the bi-metal joints will be recovered from the old containers by cutting the welds. All the ConFlat joints and connections will also be recovered.

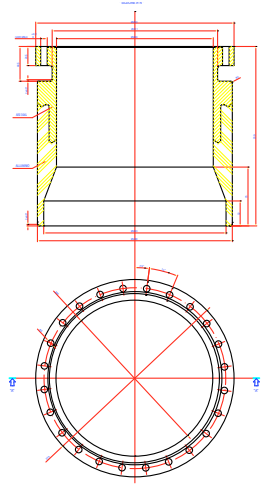

**Figure 18**: Technical drawing of a bi-metal joint.

### 2.4 The thermal shields.

The cold vessels are enclosed inside a common heat exchanger (thermal shield) into which nitrogen in two-phase condition is circulated (Figure 19). The mass ratio between the liquid and the gas is kept to less than 5:1 in order to ensure temperature uniformity all along the shield.

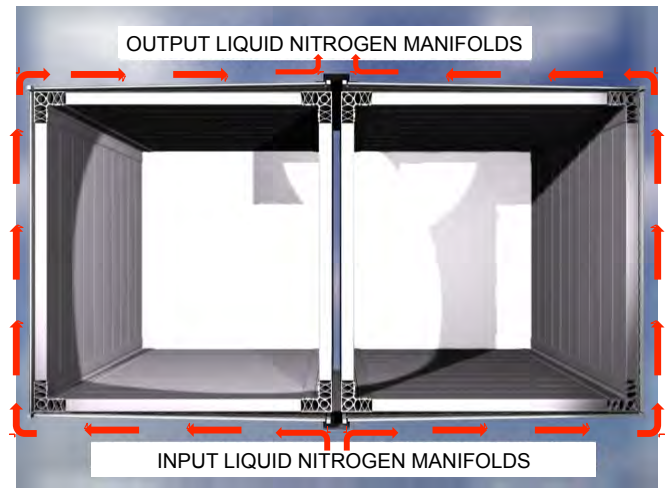

**Figure 19:** Rendering of the Nitrogen shields for the T600, Nitrogen flow is indicated with the red arrows.

This shield is an essential component of the detector since it minimizes the heat load to the bulk of the sensitive LAr volume allowing to: (1) suppress any risk of LAr boiling that could be detrimental for the HV system; (2) establish very good temperature uniformity ($\Delta T < 1$ K), through LAr de-stratification as required for uniformity of electron drift velocity and LAr purity; (3) reduce the internal temperature gradients during the cool down making it faster and therefore minimizing the critical period with the argon purifier switched-off, important to obtain a good initial LAr purity.

With an appropriate design, namely minimal frictional flow impedance, an optimal slope and elevation differences, the thermal shield can be operated in a simplified way with liquid Nitrogen circulation driven by gravity. The same design applies also to the T150 detector.

### 2.5    The new thermal insulations.

A purely passive insulation is chosen for the installation at CERN. Taking into account the size and geometry of the internal cold vessels, this solution seems the most adequate in term of cost effectiveness, reliability and safety. There are several materials (powders, foams, aerogel), of common application, that could be used for the thermal insulation layer. Typical thermal conductivity at room temperature, for these materials not under vacuum, ranges from 20 mW/m/K to 40 mW/m/K. Since the overall thermal efficiency strongly depends on the assembly details, an industrial solution will be adopted, with proven technical design, adequate choice of materials and well established assembly and test procedures.

Membrane tanks represent a possible suited choice. This technique has been developed for 50 years and is widely used for large industrial storage vessels and ships for liquefied natural gas [9] (Figure 20). This is the reference choice for the large (20 kton) cryostats of the LBNE project in USA.

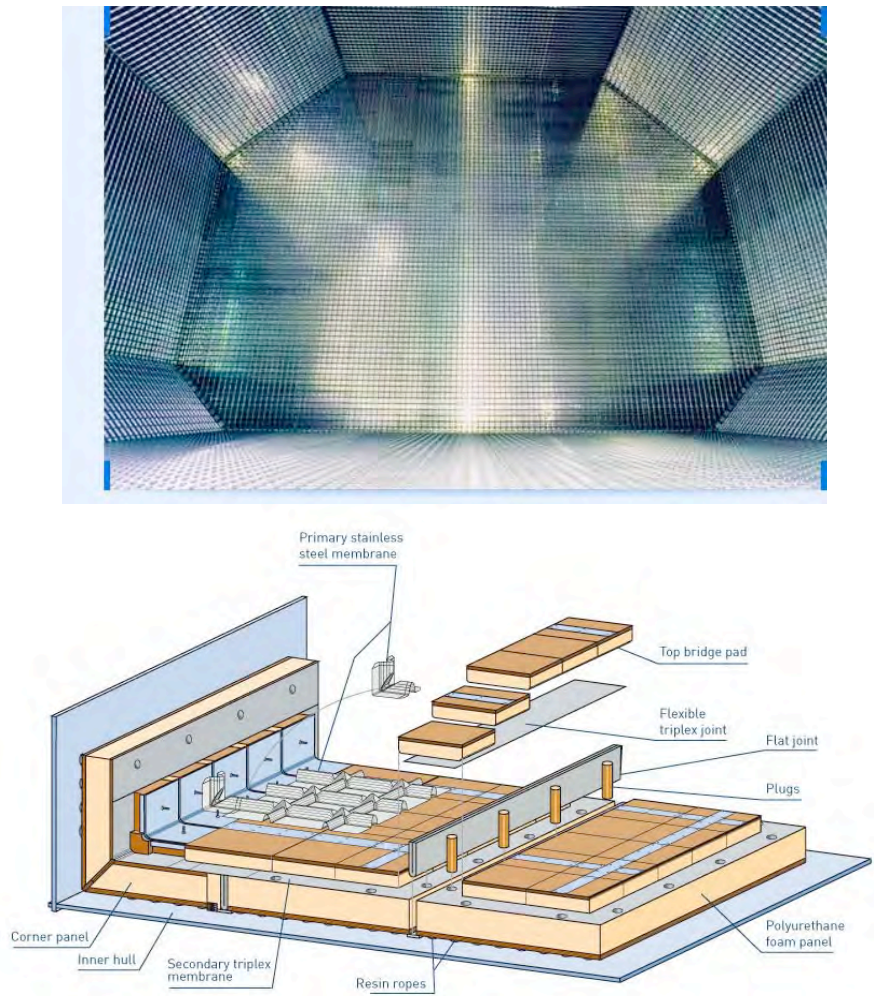

**Figure 20**: Example of application of the membrane tank technology. Top: interior of a liquefied natural gas container in a cargo ship. Bottom: scheme of construction of a membrane tank. All pictures are taken from the Gaztransport & Technigaz site: www.gtt.fr.

The details will be finalized after consultancy with the companies that will take the responsibility of design and construction. The internal stainless steel primary membrane will act as secondary liquid argon containment vessel. In case of a spill of LAr from the internal cold vessels, the argon will be collected inside the thermal insulation layer that is at the same temperature. Therefore a very limited flash evaporation will occur. This is the same principle of operation of the detector in LNGS.

An insulation thickness of 1 m will be used for the bottom and lateral sides. For the top side, to preserve the present detector layout (cables length, chimneys, HV feedthrough, LAr and $LN_2$ transfer lines) a maximum thickness of about 0.5 m will be required. An effective thermal conductivity of 30 mW/m/K can be assumed (e.g. with the configuration shown in Figure 21), resulting in a heat loss through the insulation of ~ 6.6 kW. The additional heat loss from the 20 feet supporting the cold vessel structure is estimated to be less than 2 kW. All the external contributions (cables, pumps, transfer lines, etc.) can be accounted for a value not exceeding 5.4 kW leading to a total heat load of about 14 kW, a major improvement with respect to the present 30 kW consumption rate al LNGS.

In the case of the T150, the expected heat loss will be ~ 3.5 kW through the insulation, and 0.5 kW from the feet, while the external contributions will not exceed 2 kW for a total heat load of about 6 kW.

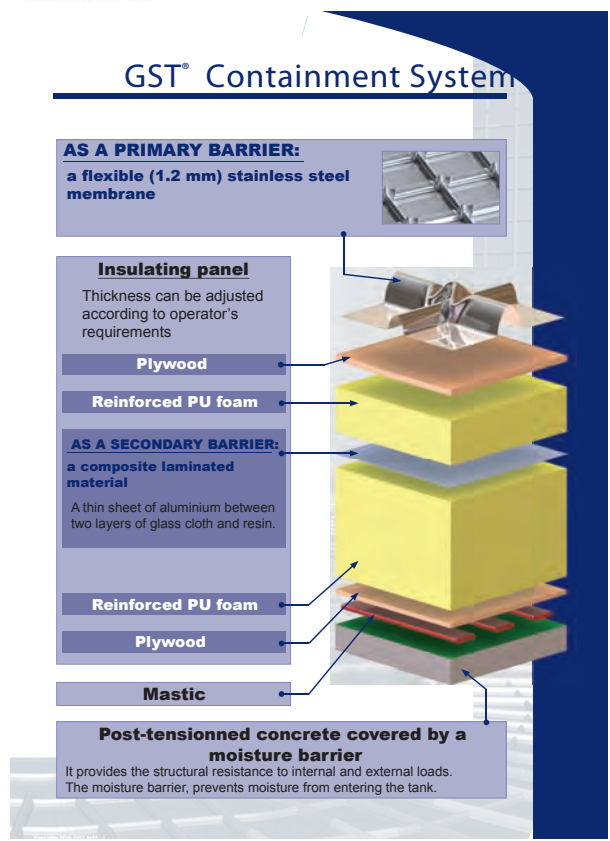

**Figure 21**: Principle of the membrane tank. Taken from Gaztransport & Technigaz site: www.gtt.fr.

### 2.6    LAr purification systems and cryogenic plant

Argon purification is one of the most fundamental issues for the successful development and operation of the LAr TPC technology. The solution that was developed by ICARUS over about 20 years of R&D is essentially based on three key elements: (1) Use of commercial filters, carefully selected among the many available on the market; (2) Ultra High Vacuum techniques and disciplines; (3) Continuous argon purification through recirculation both in the liquid and in the gas phase, which is essential for the detector operation over long time periods.

In terms of residual contamination of Oxygen equivalent (electron capture cross-section) a level of purity of the order of few tens of ppt (parts per trillion) is currently reached (Figure 10) [8]. In dedicated smaller scale tests in the lab values of few ppt were reached. These values correspond to LAr free electrons lifetimes of 6 ms and > 20 ms respectively [10].

Each T600 module is equipped with: (1) two GAr recirculation units, placed at the two ends of the module, each one designed for a maximum flow rate of 25 $Nm^3/hr$; (2) one LAr recirculation unit working at a nominal rate of 2 liquid $m^3/hr$; LAr is re-injected, after purification by the filters, on the opposite side (about 20 m apart) of the vessel; (3) one main purification unit composed by 4 hydro/oxysorb filters in parallel, which is used both for the LAr recirculation and for purification during the filling phase.

The T150 detector will be equipped with a system similar to the half of the T600 volume, with the exception that only one GAr recirculation unit will be installed. The overall purification capability will be therefore superior to the one presently available in the T600.

The GAr recirculation system also performs as a pressure controller. Heat entering from the top through the cables, flanges, etc., is only partially dumped by the thermal shield and it produces a surface evaporation of the LAr, which is re-condensed and purified by recirculation systems and then injected back into the vessels.

GAr re-condensation power is provided by $LN_2$ forced flow into a heat exchanger placed at the entrance of the GAr recirculation systems. The LAr recirculation system is cooled by an extension of the same forced $LN_2$ cooling circuit (for the argon pump, the filters and the return line) to prevent gas production that would significantly reduce the mass flow and which will be detrimental for the HV system.

The principal cooling power is provided by circulation into the thermal shield of liquid and gas nitrogen mixture. Forced circulation with centrifugal pump is adopted as standard operating mode (Figure 22). However the present installation allows the operation of the cooling system in pure passive mode by means of $LN_2$ gravity circulation through the thermal shields and additional gas condensers for pressure control. This ensures the cryogenic operation of ICARUS, without the argon purifiers, also in case of long blackouts or $LN_2$ pumps failure. The collaboration is considering the possibility to upgrade the system by including the purification into the additional argon condensers circuits operating ICARUS permanently with a passive LN$_2$ circulation. This would require a 2 months test-run before the decommissioning at LNGS in order to validate the solutions adopted.

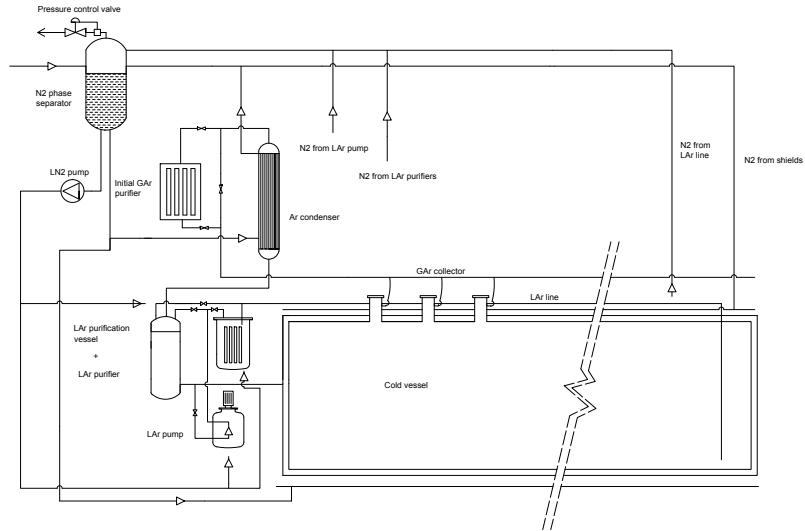

**Figure 22**: Nitrogen circuit P&I scheme illustrating the cooling by gravity circulation.

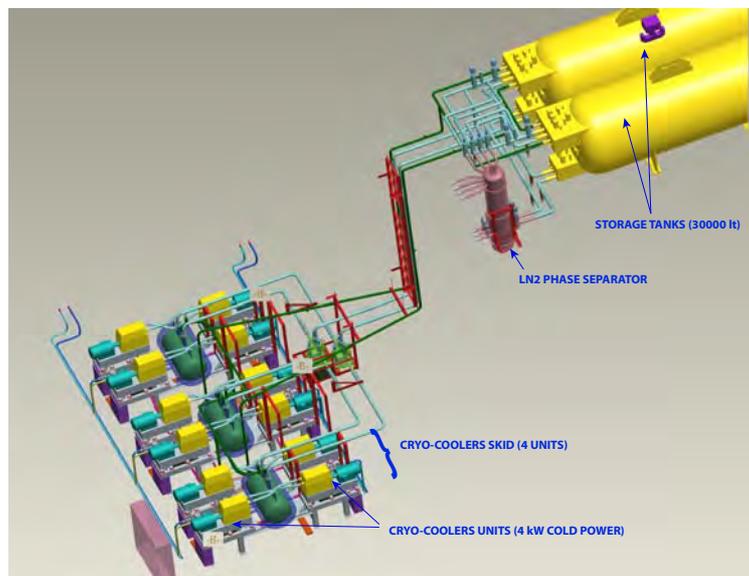

**Figure 23**: Rendered view of the Nitrogen re-condensation system based on the Stirling cryo-coolers as it is in the LNGS installation.

The LN$_2$ maintains the requested uniform temperature of the LAr. The cooling capacity needed to re-condense the gaseous Nitrogen generated by the entire system is provided by Stirling cryo-coolers producing up to 4 kW of cold power each at 80 K. The consumption rate expected for the CERN installation is of about 14 kW for the T600 and of about 6 kW for the T150. Therefore the global LN$_2$ consumption will be handled by a fraction of the 12 Stirling units presently available at LNGS (Figure 23). In LNGS the cryo-coolers are located at the ground level; at CERN these units will be more conveniently placed at the same level of the storage tanks or above.

### 2.7    *Process control and handling of the cryogenic plant.*

The LAr-TPC's plant (Argon purification and Nitrogen circulation systems, manufactured by Air Liquide) and the Stirling cryo-coolers system are both equipped with local and independent supervisors and control systems, based on widely employed industrial standards (Allen-Bradley for the T600/T150 cryogenic plant and Hitachi for each cryo-cooler, in order to have a more reliable decentralized and diffused control system for the Stirling plant). A high level of redundancy of both the control systems and all the related equipment (pumps, valves, sensors, UPS, etc.) is installed with automatic switching to guarantee the maximum operational continuity. A common interface links with the SCADA remote supervisor and control system (Figure 24). This system, that also records historical data of all the relevant parameters and events, is interfaced with the general monitoring system of LNGS. It issues alarms and automatic notifications. The same system will be interfaced with the general monitoring supervising system of CERN.

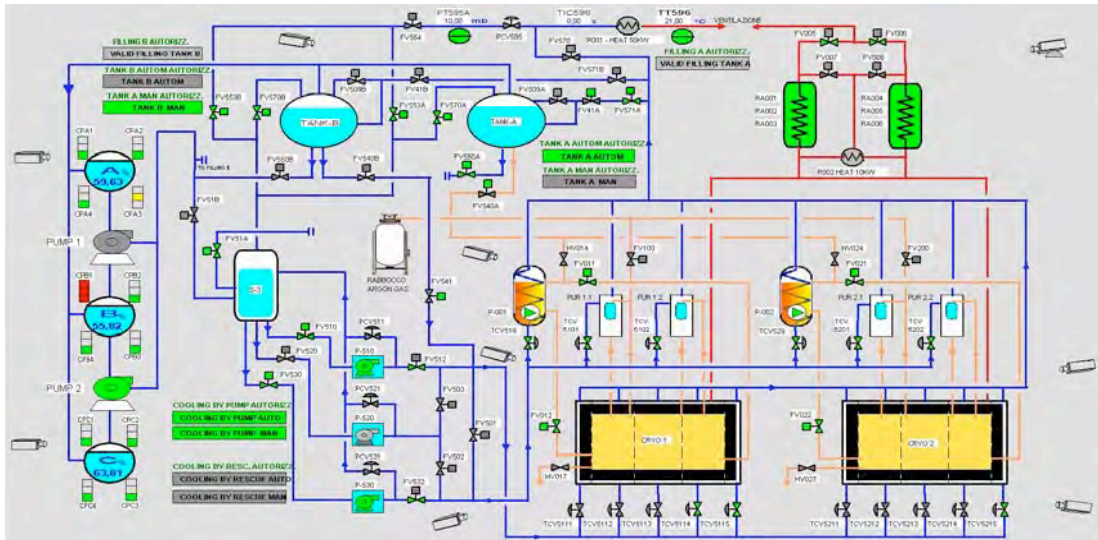

**Figure 24**: Principal console of the SCADA supervisor system.

### 2.8    *Construction of the T150 internal detector.*

The present design of the T600 is extended [11] to the basic structure of the internal detector of the T150 module. The module contains a high precision, high stability stainless steel structure independent of the container that supports two wire chambers, with three read-out planes each, the field shaping electrodes and one cathode, separating the two 1.5 m drift regions. This gives the possibility to overcome the problems connected with the relative shrinking between the aluminium vessel and the stainless steel frame of the wire chambers. The PMT system for scintillation light detection and all the Slow Control sensors (level meters, position meters, temperature probes etc…) are anchored to the sustaining structure. Most of them are positioned behind the wire planes in the inter-space of about 30 cm between the last wire plane (Collection) and the LAr container wall.

Most of the solutions already successfully adopted for the T600 at LNGS will be used: (1) simplified wire chamber structure, with three parallel wire planes at 0° and ± 60° from horizontal, 3 mm pitch and 3 mm plane separation; (2) AISI 304L stainless steel sense wires with 0.15 mm diameter; (3) a similar choice of the geometry of the field shaping electrodes made of 35 mm diameter tubes with 50 mm pitch; (4) a similar cathode structure.

Following the design of the T600 detector, the electric field is 500 V/cm for a drift path of 1.5 m. The applied HV is -75 kV. The cathode is built up by an array of panels made of punched stainless-steel sheets with a 58% optical transparency between the two drift regions. The electric field in each drift volume is made uniform by means of the field shaping electrodes (race tracks), consisting of rectangular rings made of stainless-steel tubular elements connected by two welded terminals. The electric field shielding in the upper part of the detector is made by grounded metallic panels placed above the HV system and immersed under the LAr free surface. The cathode HV distribution to the race tracks are carried out by connecting four 100 MΩ resistors in parallel: the resulting 25 MΩ resistance realizes a 2.5 kV drop between one electrode and the following one.

In Table 6 the main parameters of the internal detector mechanics for T150 module are shown; a sketch of the internal detector mechanics is also shown in Figure 25.

**Table 6.** Main parameters of the internal detector mechanics for the T150 module.

| | |
|---|---|
| Number of read-out chambers | 2 |
| Number of wire planes per chamber | 3 (all read-out) |
| Wire orientation with respect to the horizontal | 0°, ±60° |
| Wire pitch (normal to the wire direction) | 3 mm |
| Wire length: | |
|     horizontal wires | 9.42 m |
|     wires at ±60° | 3.77 m |
|     wires at the borders (±60°) | 0.49 ÷ 3.77 m |
| Wire diameter | 150 mm |
| Wire nominal tension | 12 N |
| Number of wires per chamber: | |
|     horizontal wires | 1056 |
|     wires at ±60° | 2 x 2336 |
|     wires at the borders (±60°) | 2 x 960 |
| Total number of wires | 15296 |
| Maximum drift length | 1.5 m |
| Distance between race tracks | 50 mm |
| Number of race tracks per sensitive volume | 58 |
| Imaging volume (mass) | 85.1 m$^3$ (119 t) |
|     Length | 8.98 m |
|     Width | 3 m |
|     Height | 3.16 m |

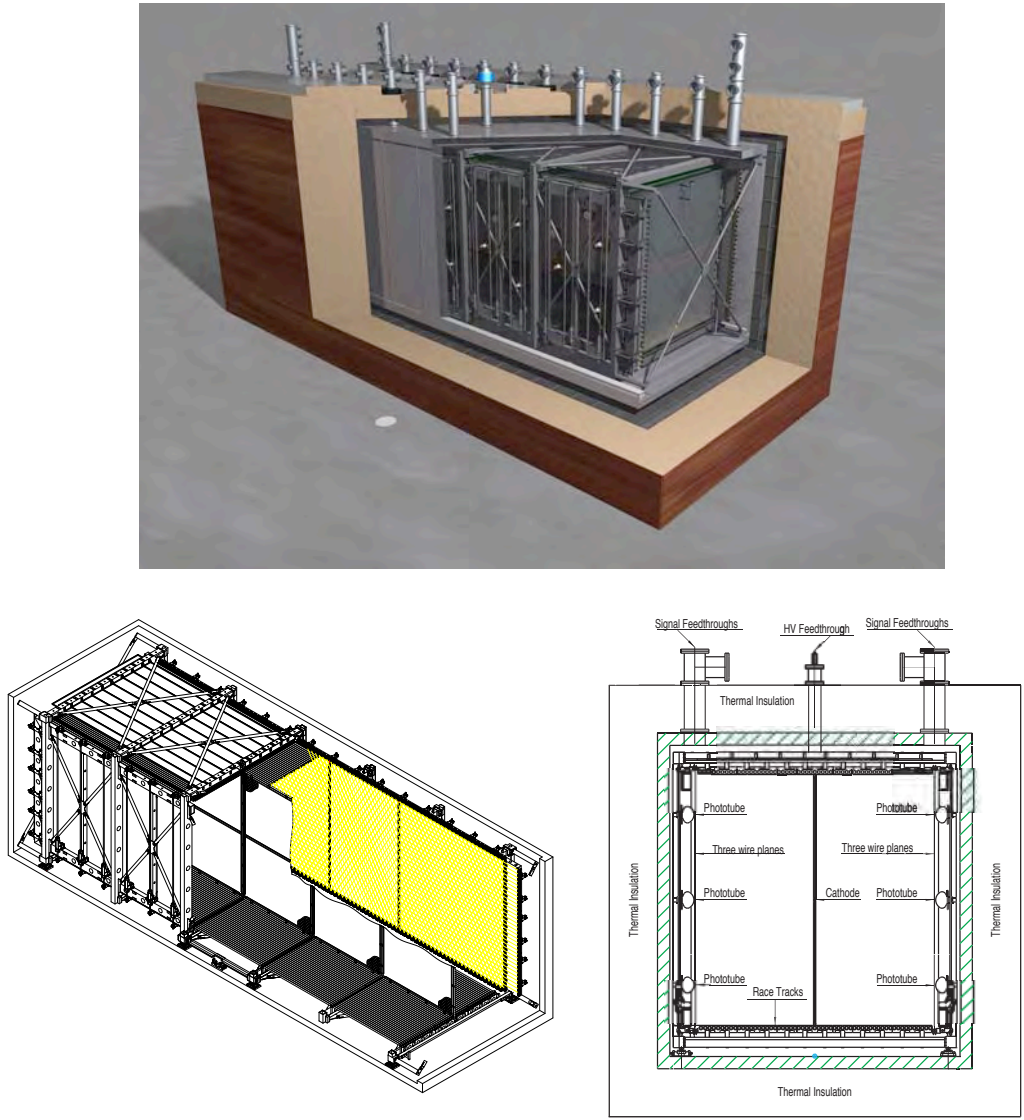

**Fig. 25:** T150 detector views.

Most of the components presently used for the T600 wire chambers (wire holders, combs, spacers, etc.) and for the drift HV system (voltage dividers, feed-through, etc.) can be used without modification. The tooling for the production of the wires is already available, being the same employed for the T600, hence the wire production for the T150 could start as soon as the experiment is approved. Similarly, the internal detector mechanics component production is to be started by the CINEL Company, which already has all the technical drawings. Then, the assembly of the internal detector is performed at CERN inside the T150 cryostat. A clean room directly connected to the front side of the LAr vessel is in a conditioned atmosphere. After the support structure assembly, the installation of the ≈ 15,000 wires will be carried out.

### 2.9    Upgrade of PMT systems.

Following the T600, the 128 nm prompt scintillation light is exploited to provide an absolute time ($t_0$) for the event and an absolute scale in the electron drift direction.

The most common approach to observe the VUV scintillation light with PMT's is to coat, by evaporation (Figure 26), the photocathode external window with a tetra phenyl butadiene (TPB) wave-shifter, which has been shown to work in liquid argon with good efficiency and stability. About 100 μg cm$^{-2}$ of TPB are enough to act as a wavelength shifter in the visible region [12]. PMTs provided with sand blasted glass windows, allows increasing the compound adhesion. The solution presently adopted in the T600 is the use of large surface ETL 9357-FLA™ PMTs: a 12-stage dynode PMT with hemispherical 200 mm (8 in.) diameter glass window, manufactured to work at cryogenics temperatures [13].

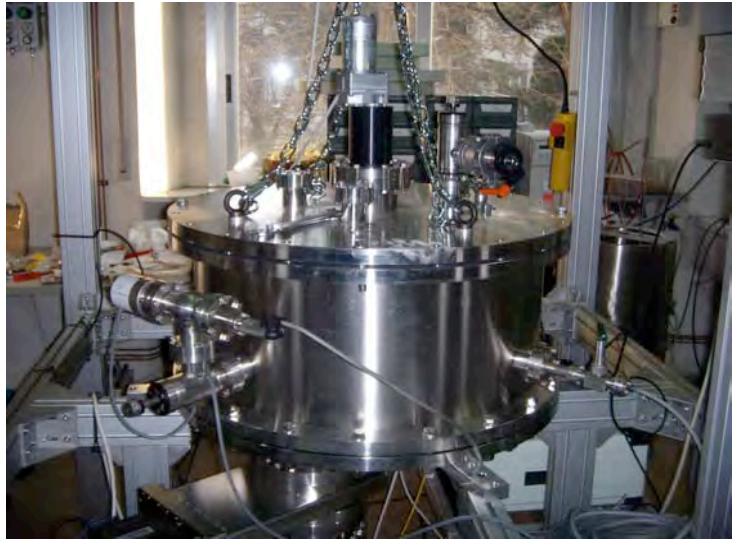

**Figure 26.** Vacuum evaporation system for wavelength shifter deposition.

The light detection coverage in the T600 exploits the present PMT scheme by deploying for each chamber a total of 27 PMTs behind each wire chamber (Figure 27). For this layout the evaluated efficiencies for light collection in the CERN neutrino beam are enough to collect almost all the events. Old PMT's will be replaced with new devices (8 in. Hamamatsu R5912MOD™ PMT), offering better performance, such as a higher and more uniform quantum efficiency on the sensitive surface (Table 7).

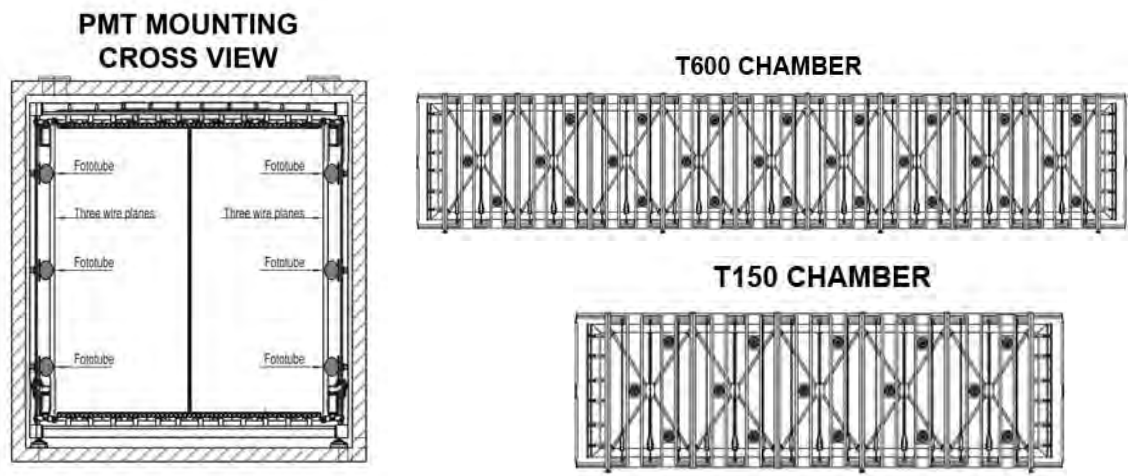

**Figure 27.** PMT installation layout behind each T600 and T150 TPC wire chamber. The distance between PMT's in each row is 2 m. The rows separation is 1 m. In the first T300 module only the middle row is presently installed.

**Table 7.** Main characteristics of Hamamatsu R5912-MOD PMTs

| Cathode size and type | 190 mm, Bialkali on a Pt layer |
|---|---|
| Spectral resp. and QE | 300-600 nm, 17% (420 nm) |
| HV and Gain (typ) | 1500 V, $10^7$ |
| Risetime, TTS (FWHM) | 3.6 ns, 2.4 ns |
| SER peak-to-valley | 2.5 |
| # stages | 10 |

The detection efficiency of the low energy neutrino events could be increased up to a factor of 2 after the proposed modification of the PMT light collection, which is schematically illustrated in Figure 28. The light detection efficiency is expected to be significantly improved with the use of a Teflon light guide surrounding the PMT photocathode with the height of about 8 cm. The inner surfaces of the light guides could be coated with the TPB WLS to convert into blue light, which is diffusively reflected from the Teflon and which partially reaches the PMT window. Simulations and R&D will be performed to optimize the shape of the light guides in order to achieve the maximum light collection. The installation of the light guides is well adapted to present PMT structure, it requires a minimum intervention and it does not need any modification of the internal detector.

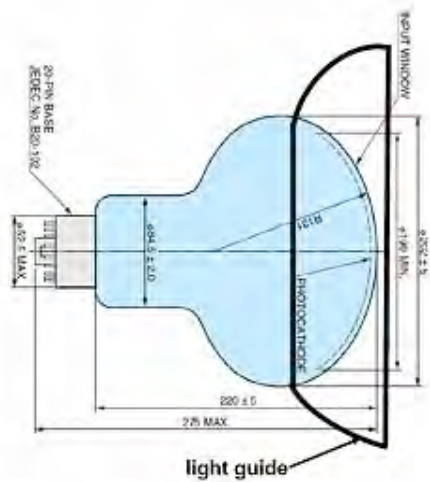

**Figure 28.** Schematic view of Hamamatsu R-5912MOD PMT with the proposed light guide.

The voltage divider designed to operate at low temperature is welded directly to the PMT output leads. It will be optimized to provide fast trigger, light amplitude measuring and monitoring.

Following the present T600 layout, the PMT system makes use of analogue signal adders to provide the data acquisition with a trigger signal related to the scintillation light and hence to the energy deposited in LAr. Waveform digitizers with circular buffering may also be used to sample and acquire each PMT signal with the same DAQ layout adopted for the wire signals. The identification of the time sequence of ionizing events extracted from the PMT pattern is useful for the background identification.

The PMTs layout and electronics of the T150 detector will be similar to that of T600, deploying about 15 PMTs behind each wire chamber. Alternative solutions for the light detection, such as silicon photomultiplier (SIPMT) [14], large-area avalanche photodiodes (LAAPD) or light guide detector [15] will be evaluated as an option for the T150 detector.

### 2.10    Readout electronics

The ICARUS T600 electronics is based on an analogue front-end amplifier followed by a multiplexed 10-bit AD converter and by a digital VME module that performs local storage and data compression [11,16]. It is installed on the top of the cryostats, connected to the TPC wires through custom made feedthrough mounted in UHV grade flanges. The overall architecture based on VME standard and performance are perfectly adequate for the proposed experiment, nevertheless possible improvements could always be implemented, being the whole electronic chain fully accessible.

The new T150 detector requires the implementation of 15296 new electronics channels, based on the same analogue front-end amplifier used in the T600. The only change is the adoption of a smaller package for the BiCMOS custom amplifier, dual channel, part of the front end, which is already available. The same architecture of T600 is adopted also for the digital part, with some changes concerning the implementation with more up-to-date components.

For T600 detector a very reliable and cost effective flange has been developed, which allows the connections of 18 cables, twisted pairs, and each conveying 32 signals from wire chamber to external electronics. For T150, the external side of the flange is a sort of backplane, that support both analogue and digital electronics, with a compact design to house the full electronics on the flanges (Figure 29). The overall cost will be drastically reduced.

Performance of the read-out system can be improved replacing the VME (8 - 10 MB/s) and the sequential order single board access mode inherent to the shared bus architecture, with a modern switched I/O, as PCI Express standard for instance, allowing the parallelization of the data flows. In addition such I/O transaction can be carried over low cost optical Gigabit/s serial links. A possible solution already in development is described (Figure 30). Presently commercially available serial links are under test, but one can seriously consider implementing the White Rabbit (WR) serial high-speed connection. The WR project is a multi-laboratory and multi-company effort to bring together the best of the data transfer and timing worlds in a completely open design. It takes advantage of the latest developments for improving timing over Ethernet, such as IEEE 1588 (Precision Time Protocol) and Synchronous Ethernet. The WR architecture based on Gigabit Ethernet will perfectly match the event building structure.

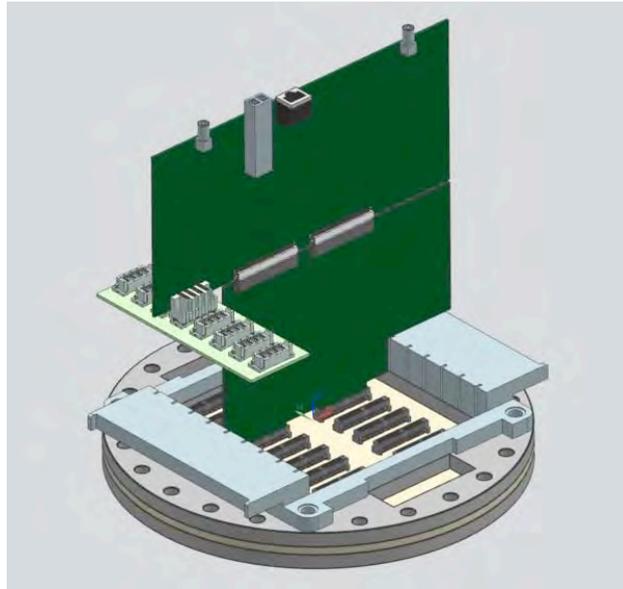

**Figure 29.** Rendering of the flange and electronics boards under development.

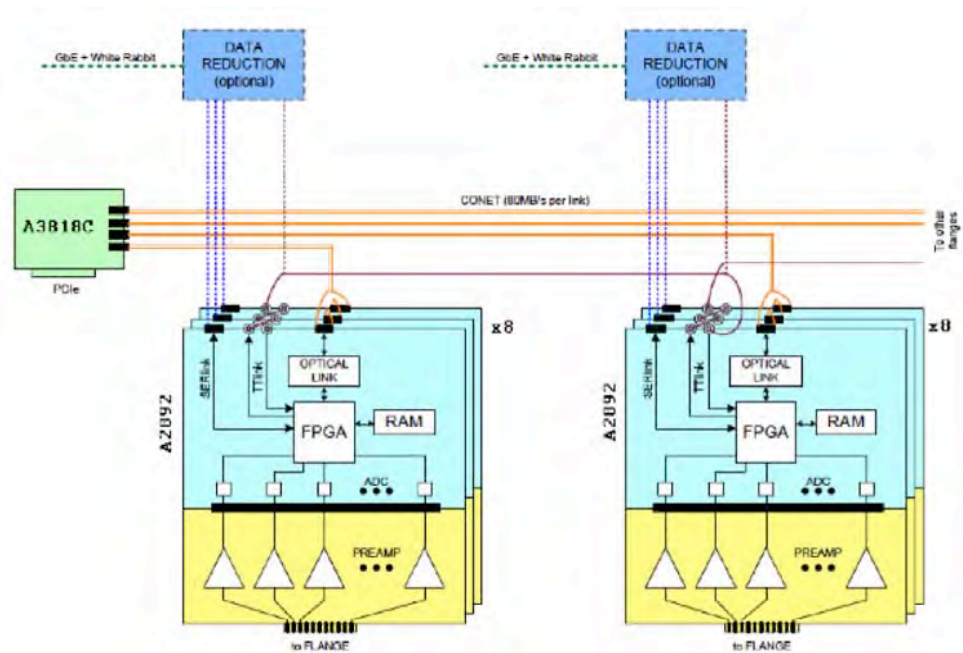

**Figure 30.** Block diagram of electronics housed onto the flanges.

### 2.11    *Additional instrumentation for detector slow control*

In order to monitor and control the operation and behaviour of the T600 and T150 detectors in the critical phases of the vacuum, cooling and filling with LAr, a series of sensors with the corresponding read-out electronics are fore-seen. The T150 Slow Control system will be a copy of the T600 one at LNGS. As for one T600 module, the following sensors will be installed inside the T150 cryostat: sixteen cylindrical capacitor liquid argon (LAr) level meters, placed at the four corners of the cryostat, to monitor the LAr filling of the internal vessel; twenty resistors to monitor the LAr level at regime; eight cylindrical capacitor position meters to measure the inward movement of the vessel when is being pumped before filling it and the outward movement when it is filled with LAr. Thirty platinum resistors (Pt1000) for the internal temperature measurement, arranged in six groups by five each, one in each corner and two in central position, distributed along the height of the cryostat.

Moreover, the T600 is equipped with 12 LAr purity monitors, consisting of small double-gridded drift chambers, installed outside the LAr active volume at different heights. Bunches of electrons, extracted from the cathode via photoelectric effect, are collected at the anode, allowing the measurement of their attenuation along the drift path. This method provides an on-line and local estimate of the electron lifetime, especially during the first phases of the LAr purification. The electron lifetime measurement during the run is instead based on the analysis of muon events, by studying the attenuation of the signal amplitude as a function of the electron drift distance.

# 3 The NESSiE muon spectrometers

The need for a detector to identify the charge of the muon produced by the incoming neutrino as well to perform an extensive measurement of the CC events, i.e. the determination of muon momenta, is reported in the NESSiE proposal [3] and as referenced in Section 1. Here technical aspects are farther depicted with full consideration for a coherent design with the two LAr-TPC detectors.

We recall that some important practical constraints were assumed in order to draft the report on a conservative, manageable basis, and to maintain the proposed detector sustainable in terms of time-scale and cost. Well known technologies were considered as well as re-using parts of existing detectors (should they become available; if not that would imply an increase of the costs but no additional delay). Moreover the experience acquired by a part of the involved people in the construction, assembling and maintenance of the present running OPERA spectrometers will be largely exploited.

The momentum and charge state measurements of muons in a wide energy range, from few hundred MeV to several GeV, over more than 50 m$^2$ surface, is an extremely challenging task if constrained by a 10 (and not 100) millions CHF budget for construction and installation. Running costs have to be kept at low level, too.

This part of the experiment is identified throughout the proposal with the acronym NESSiE (Neutrino Experiment with SpectrometerS in Europe).

The main purpose of a spectrometer placed downstream of the target section is to provide charge and momentum reconstruction for muons escaping LAr detection. The two constraints of measuring both charges and momenta over a rather large energy interval are somehow in contradiction whether the case of a realistic, conservative, relatively inexpensive apparatus is taken into account. The problem has been solved by optimizing the thickness of a multi-layer iron magnet placed downstream of a magnet in air. Therefore low momentum muons, which would not cross a sufficient number of iron-layers to determine their curvature, are measured by the Air-Magnet. Instead muons with higher momenta are well measured by crossing many iron-layers. Figure 31 depicts that result for the charge identification.

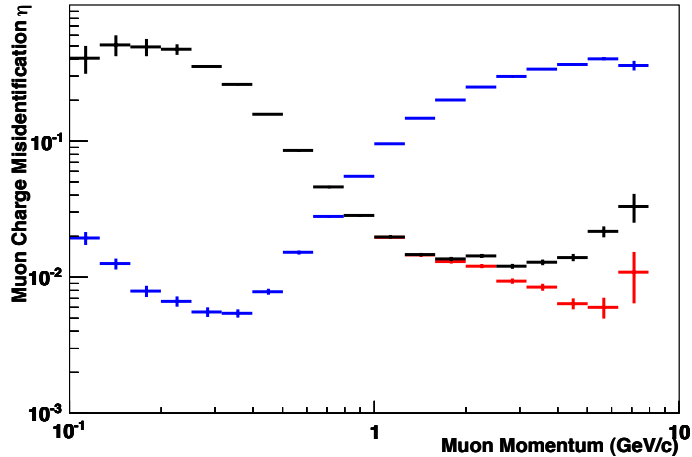

**Figure 31.** The charge mis-identification percentage including all the selection, efficiency and reconstruction procedures by the NESSiE system. The blue dots correspond to the measure performed by the magnetic field in air, the red (black) dots correspond to the magnetic field in iron with the two (one) arms.

### 3.1    The Iron Magnets

The mechanical design of the Far site Iron-Magnet is well advanced, some details being given in Figure 32. The magnet is basically built assembling iron slabs of 5860 mm height, 1246 mm wide and 50 mm thick: 294 slabs are needed to complete the magnet. The height of the Far site experimental hall is fixed to 8 m by the need of lifting up the iron slabs with the crane and then pulling them down into the NESSiE pit (see later).

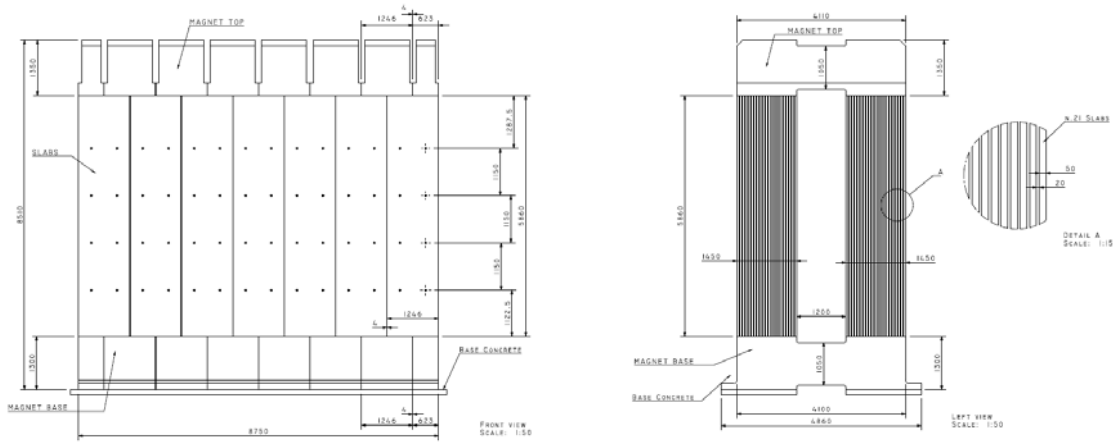

**Figure 32.** Transverse and longitudinal views of the FAR site iron magnet.

If the assembly will proceed according to the OPERA scheme, the individual iron slabs, once in the pit, will be assembled in planes using first the magnet base and a temporary support structure placed between the two arms. The construction will start from the innermost planes of the two arms. As the magnet is also instrumented by RPC detector elements, the orthogonal readout strip planes and the extra insulating material to prevent HV discharges toward the iron will be sandwiched between consecutive iron walls. Once the magnet top will be put in place the mechanical structure will be self substaining. At this stage the magnet could then be moved into its final position. The total iron mass will amount to 1515 t. The air pad technology, successfully used for the displacement of large portions of the CMS detector, could be applied. For illus-tration purposes, a test load of 1000 t being moved using a system composed of 4 high pressure air pads, is shown in Figure 33. The movement is possible also in the presence of a small slope, up to a few percent.

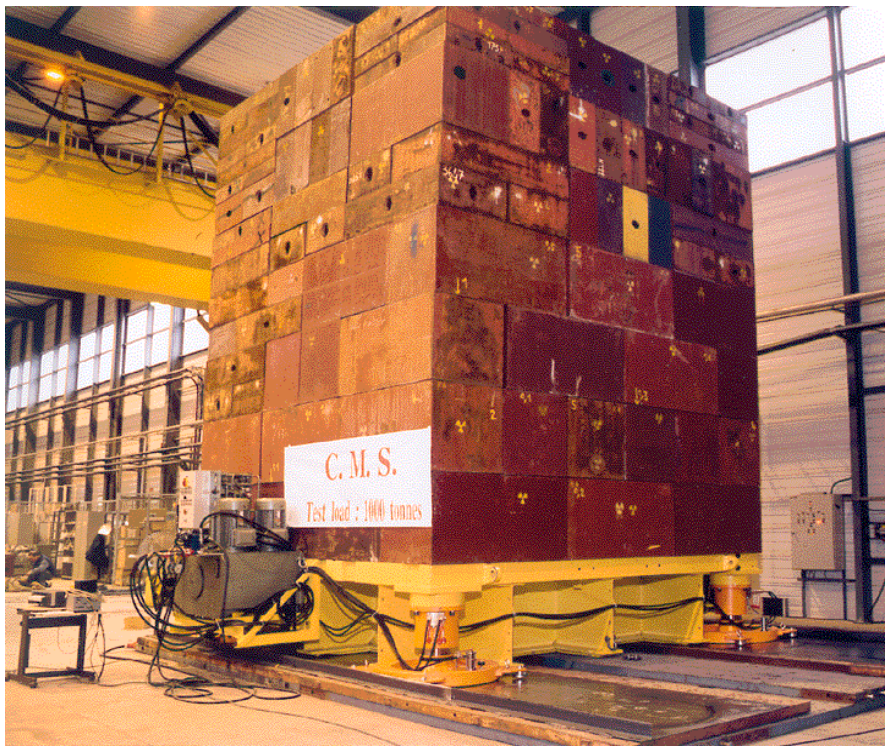

**Figure 33.** 1000 t test load being moved on a system of 4 high pressure air pads.

To illustrate in more detail the assembly scheme the example of the OPERA magnet will be followed. The initial supporting structure in shown in green and orange in Figure 34. Once the assembly is completed this area correspond to the gap between the magnet arms. The array of long horizontal rods ensures in this phase the stability of the 5.86 m high iron slabs. Once in place each slab is secured by a system set of bolts. The iron itself is underneath the plane of copper horizontal strips clearly visible in the bottom part of Figure 34. The rectangular RPC detector elements are also visible in black. At the time the picture was taken, only the two lower row of detectors were missing. The RPC detectors are propely positioned in front of the rods with the tool depicted in Figure 35.

Based on the OPERA experience the iron "chemical" composition has to satisfy other contraints besides the requirement of a high magnetic permeability. In the presence of magnetic field significant magnetic forces must be absorbed by the structure. During the assembly the structure is exposed to a significant stress, too. This leads to the following requirements for the various contaminants present in the steel (weight fractions):

- C contamination: $< 0.080$;
- P contamination: $< 0.025$;
- S contamination: $< 0.010$;
- B contamination: $< 0.0005$;

and to the following mechanical specifications:

- miminal breaking strength: 340 N / mm² ;
- minimal yield strength: 225 N / mm²;
- elongation: > 25 %.

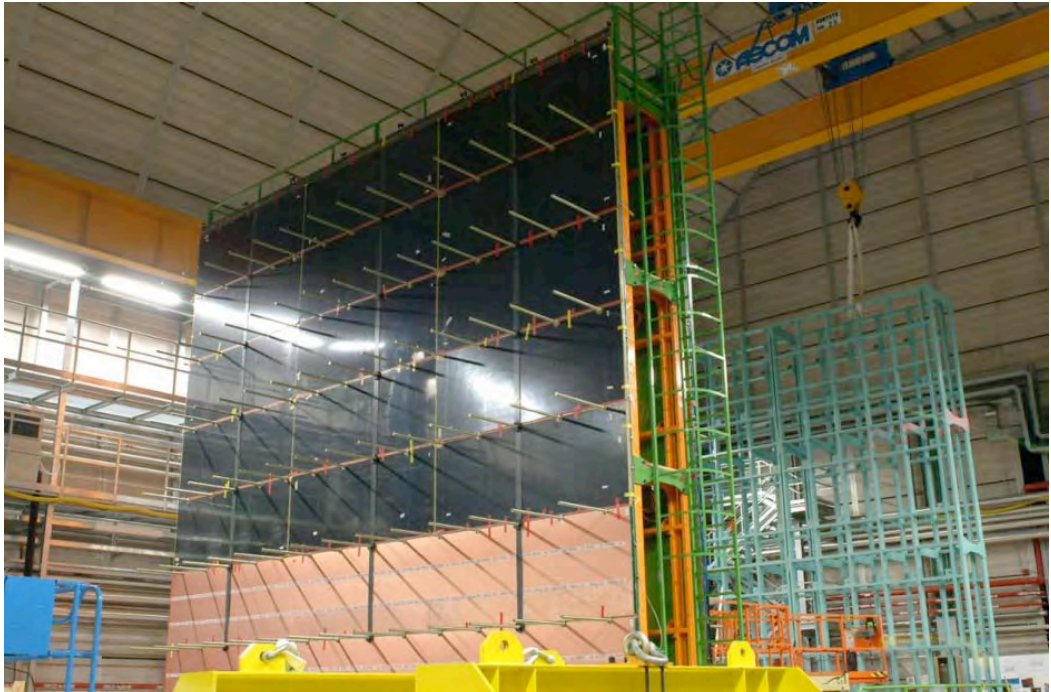

**Figure 34.** Detail of the assembly of the OPERA iron magnet.

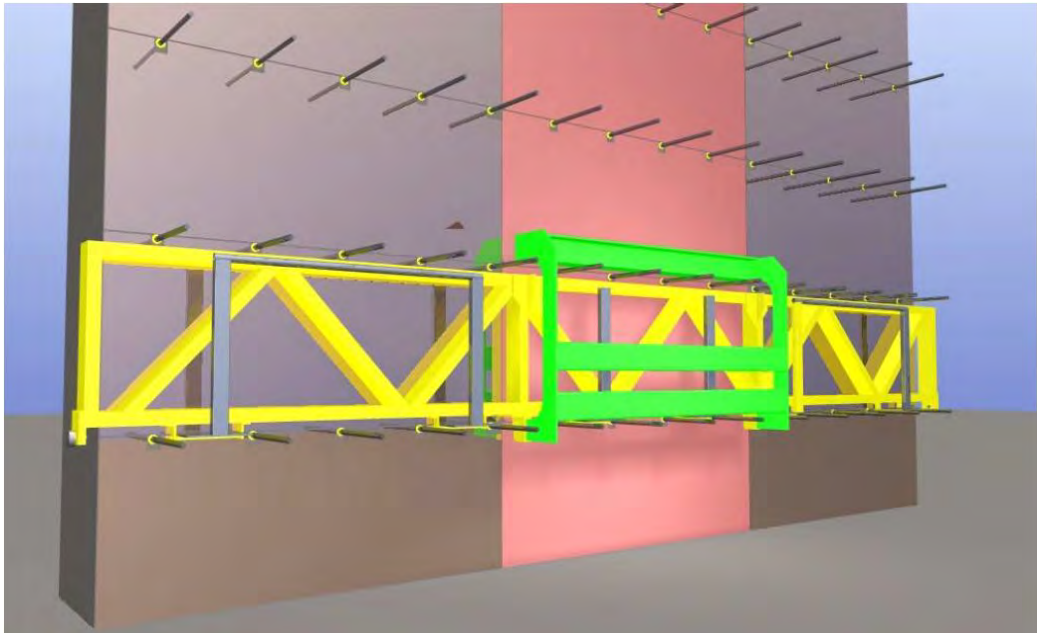

**Figure 35**. tool used to put in position 3 RPC detector elements.

Finally in order to follow the OPERA mounting scheme some tolerances on the slab dimensions must be fulfilled:

- slab height: 0.1 mm;
- planarity: 3.0 mm;
- orthogonality of the 1246 mm side w.r.t. the vertical axis: 0.1 mm;
- parallelism between the top and bottom 1246 mm sides: 0.1 mm;
- planarity of the horizontal surface of the return yokes: 0.1 mm.

The design of the air magnet part is being finalized. As this part of the magnet should also bear significant magnetic forces it should be mechanically independent from the iron magnet. The air magnet is instrumented with RPC detectors, too. Likely this magnet will be assembled outside of the pit, at the ground level. Once the assembly of the iron magnet will be completed the air magnet will be lowered down into the pit and placed into its final position.

The mechanical design of the Near site iron magnet is also well advanced, some details are given in Figure 36. The magnet is basically built assembling iron slabs of 3515 mm height, 1246 mm wide and 50 mm thick: 210 slabs are needed to complete the magnet. The height of the Near site experimental hall is fixed to 7 m by the specific requirements of the LAr detector. The total iron mass will amount to about 840 t.

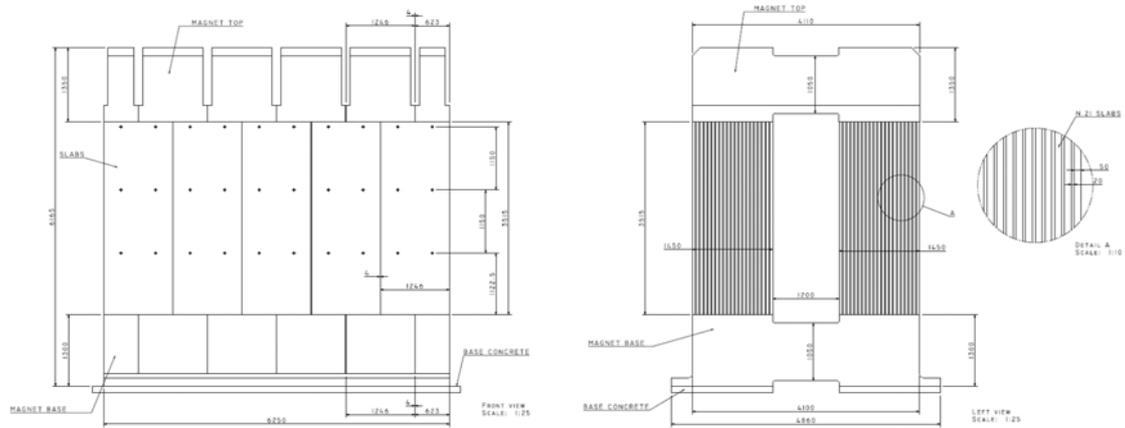

**Figure 36.** Transverse and longitudinal views of the NEAR site iron magnet.

### 3.2    *The Air Magnets*

For low momentum muons the effect of Multiple Scattering in iron is comparable to the magnetic bending and therefore the charge mis-identification increases (see Figure 37). For muon momenta less than 1 GeV/c the charge measurement can be then performed by means of a magnetic field in air. In Figure 38 the displacement expected in the bending plane is shown for muons crossing a magnetized air volume of 30 cm depth.

We choose a right-handed co-ordinate system with the origin at the interaction point and the z-axis along the beam direction, pointing towards the muon system; the y-axis points upward and x horizontally. A uniform magnetic field oriented along the y axis (the bending plane being the z-x one) is assumed. In the left plot the shift is shown as a function of the muon momentum for some values of the magnetic field in the 0.1–0.4 T range. In the right panel the spatial displacement in the bending plane estimated for muons of 0.5 GeV in a magnetic field of 0.25 T as a function of the incoming angles is plotted. The parameters of the magnet field in air are summarized in Table 8.

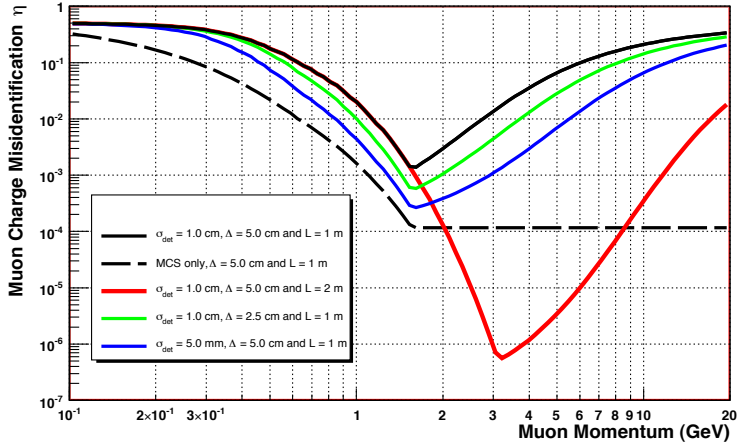

**Figure 37.** Muon charge mis-identification η as calculated for several iron-magnet configurations, with a magnetic field B = 1.5 T. Details are given in [3].

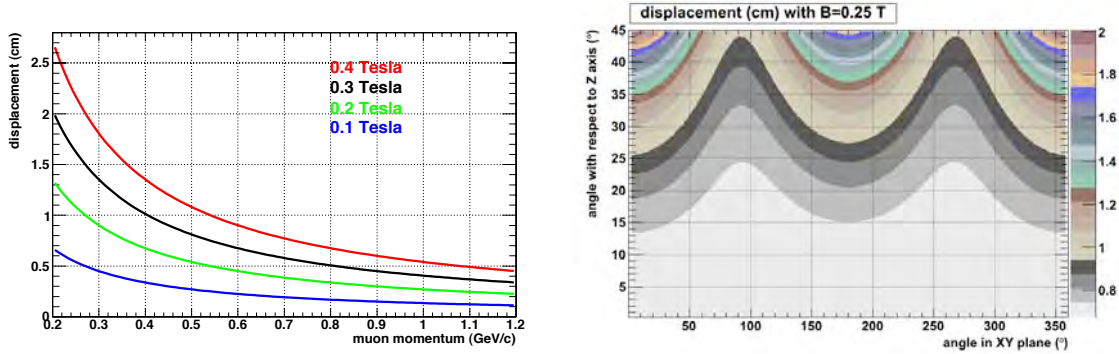

**Figure 38.** Left: Spatial displacement of muon trajectory in the bending plane as a function of their momentum after crossing an air magnetic field volume of 30 cm thickness. Different field values in the 0.1-0.4 T range are considered. The incoming muon direction is perpendicular to the detector planes. Right: the displacement of 0.5 GeV muon tracks in the x, z view (B = 0.25 T) as a function of the initial muon angle (the color bar scale is in cm).

### 3.2.1 Air Magnet Design

#### Coil description

The coil (pancake) design and construction have not only to fulfill the magnetic, structural and thermal requirements but also to take into account the necessity of having minimal material in the beam direction. For these reasons, Aluminum is the material of choice, both for the conducting cable and the supporting structure. The final coil cross section is shown in Figure 39.

The complete magnet is built using 80 coils 9 meters long in the straight parts plus two half circular bending regions for the return of the conductors, outside the beam region. The electrical and hydraulics connection services and controls are outside the beam area. Each pancake consists of two layers of aluminium hollow conductor wound first from outside to inside and the other way around for the second layer. In this way both the start and the end of each pancake are on the same side and they can be easily connected electrically, leaving access to the cooling input-output water connections in each one (Figure 40). The latter is possible since coils are electrically all in series even if they are cooled one by one independently in parallel, thus allowing to maintain the hydraulics parameters in a comfortable range.

**Table 8.** Far Air Magnet Parameters

| | |
|---|---|
| Field Intensity | B = 0.25 T |
| Electric power dissipation | P = 3.17 MW |
| Stored magnetic energy | 460 KJ |
| Inductance | 0.132 H |
| **Coil and Current** | |
| Coil structure | 80 double-layer pancakes |
| Total number of turns | N = 2x80 |
| Conductor material | Aluminum Al-99.7 |
| Conductor cross-section | 27 mm x 27 mm |
| Cooling water channel x Conductor | 1 with Ø = 14 mm |
| Conductor length | 110 m per pancake |
| Conductor weight | 180 kg per pancake |
| Current in conductor | I=2640 A |
| Current density | 4.6 A/mm$^2$ |
| Total resistance | R = 0.45 Ohm @ 20 °C |
| **Cooling** | |
| Cooling requirements | all pancakes in parallel |
| Pressure drop of cooling water | 10 bar @ 20 °C |
| Total Water Flow | 2260 l/min |
| Water Temperature drop | $\Delta T$ = 20 °C |
| **Mechanics** | |
| Overall dimension | X= 9825 mm Y=7160 mm  Z= 630 mm |
| Acceptance Dimension | X= 9000 mm Y=6143 mm Beam centered |
| Magnetic gap in air | 0.3 m |
| Average Thickness in Z | 2 x 125 mm |
| Supporting Structure Material | Aluminum |
| Total Weight | 4600 Kg |

Conductors are operated with a current density of 4.6 A/mm$^2$ thus providing the required magnetic field intensity in the inner gap.

The magnet total power consumption is 3.17 MW. To dissipate it a water cooling system operating with an input temperature T of about 18 $^0$C and a $\Delta T$ of about 20 $^0$C will be used. The total flow is 2260 l/min with a pressure drop of about 10 bars.

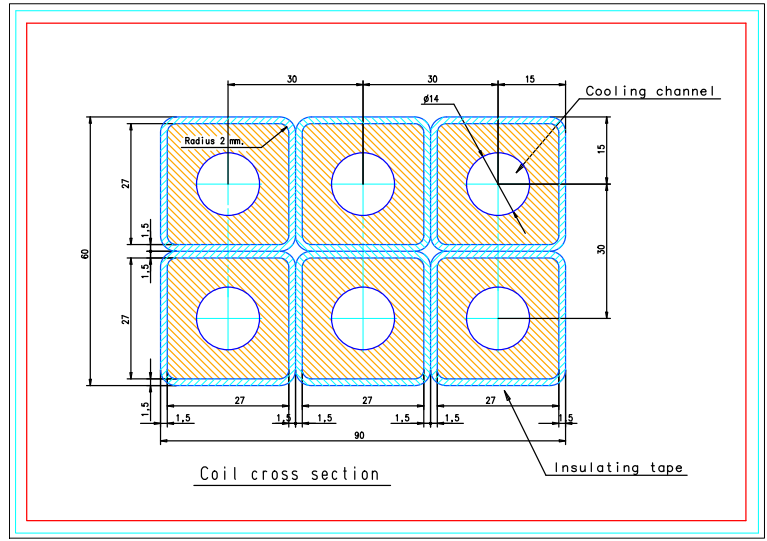

**Figure 39.** The air-magnet single coil structure.

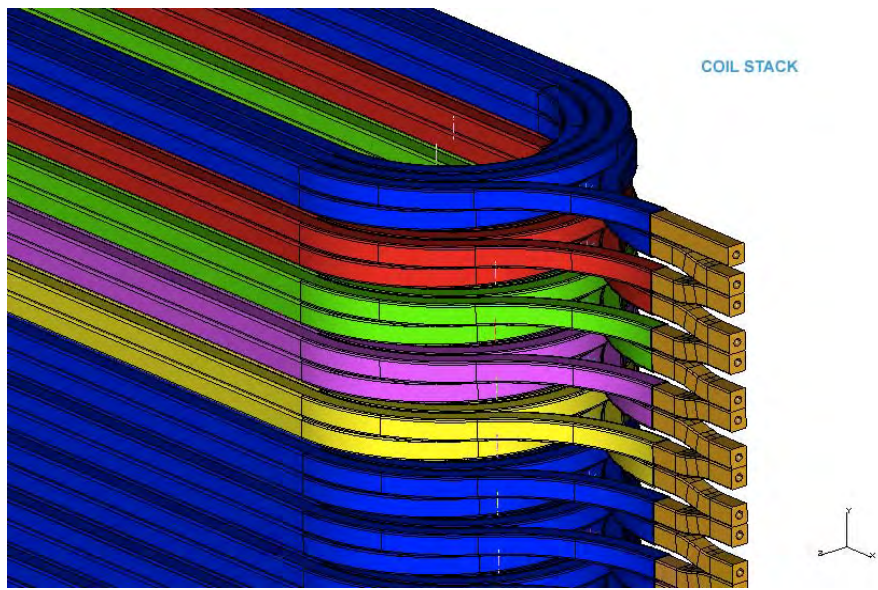

**Figure 40.** The air-magnet coil ("pancake") structure.

*Mechanical supporting structure*

A preliminary evaluation of the magnetic force was made. Each coil and the whole magnet bear the following magnetic forces: horizontally a force of about 17x10³ kg/m repels the two long side of each coil. To overcome that force the two sides will be connected serially by suitable tension bar. A back-bone (such as aluminium mechanical plate) will be coupled to the pancake lateral external sides to provide stiffness. Two stiffening plates connected to the back-bone enclose the pancakes on the top and bottom surface and provide the fixation points of the tension rods mechanically tying the two sides (Figures 41 and 42). In this way all the magnetic forces are transferred to the supporting structure.

The coils are all built the same so the final magnet results in a stack of coils packed one on top of the other and mechanically connected to give the supporting structure. An attractive force tends to compress the coils vertically.

The force is 300 kg/m in the outmost one and decreasing to zero in the central one, such that a low maximum force of about $10 \times 10^3$ kg/m results in the central part. The latter will be easily supported by the mechanical structure, which will be vertically compressed. Due to the presence of the large magnetized iron a residual force along the beam direction will attract air magnet toward the iron one. Being of the order of 200 kg/m it will be enough to connect in a few points (12) the two magnets with horizontal struts.

A detailed CAD model and a FEM analysis of the magnet and relative parts will be performed to calculate the final structure of the system and keep stresses in the necessary limits.

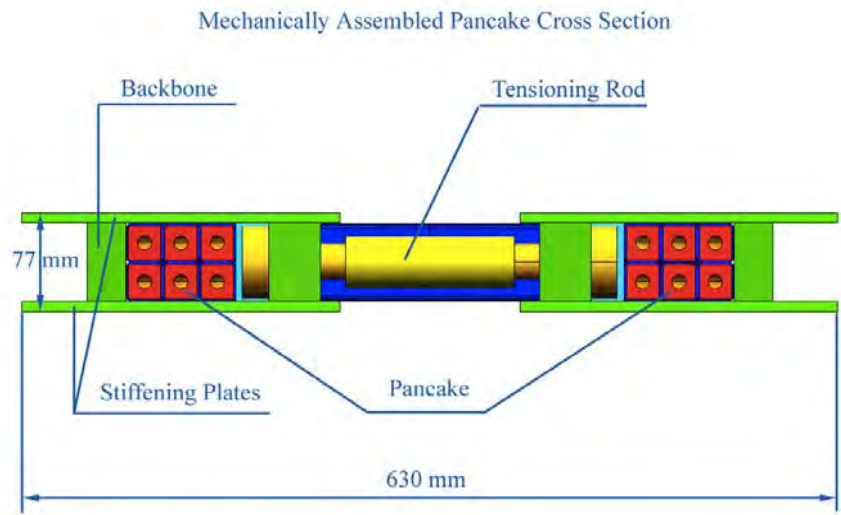

**Figure 41.** The air-magnet single "pancake" profile.

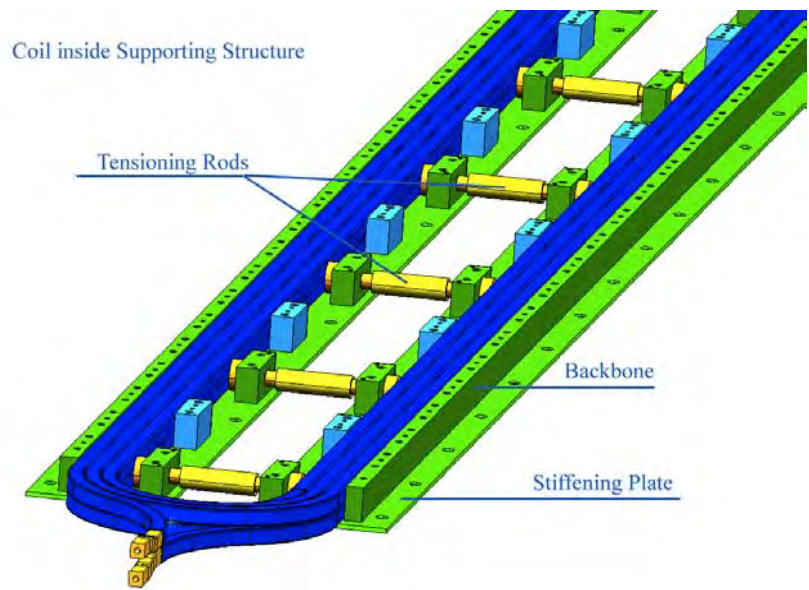

**Figure 42.** The complete assembled "pancake" for the air-magnet.

*Construction*

The high purity aluminium conductor must be custom extruded for the purpose. In the present preliminary configuration the cross section dimensions are 27x27 mm². The length of the conductor in a single pancake (Far detector) is approximately 110 m.

Nearly 10500 m are needed for the Far Detector Air Magnet and 6000 m for the Near one (built in the same way, only with reduced overall dimensions). Considering that all mechanical forces are absorbed by the supporting structure, after the curing no special structural requirement is posed on the pancake itself which will be mounted inside the supporting individual structure.

Because of the large length of the coils the best way to ensure the necessary conductor insulation is to use a pre-impregnated glass-fiber tape wound around and cured at the right temperature under vacuum.

Particular care must be used and accurate inspection must be performed in the bending area, due to the narrow bending radius. Each pancakes has its own mechanical supporting structure, contrasting the magnetic force, thus allowing to test them one by one in the final configuration, in either electrical, mechanical and cooling modes (interconnection excluded).

The fact that all the coils are identical allows a minimization of the winding and manipulating tools. Coils are also completely interchangeable (first and last one excluded) (Figure 43).

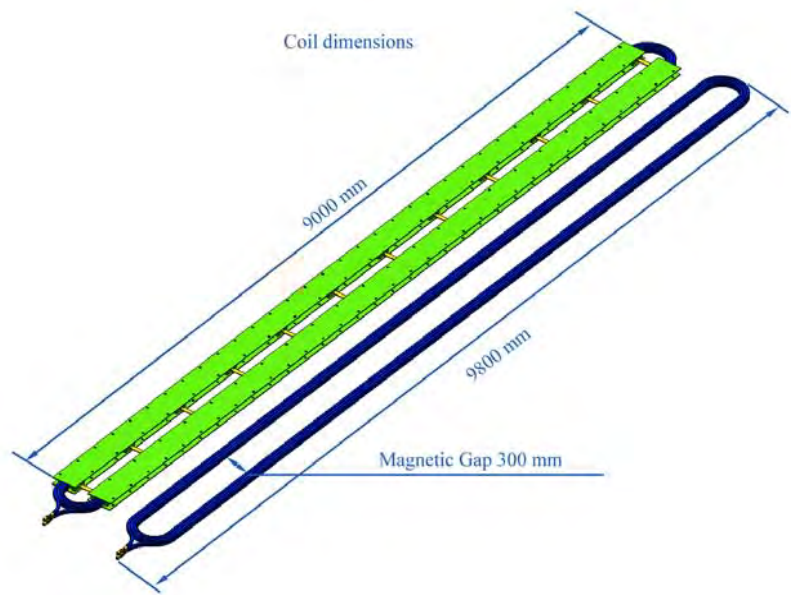

**Figure 43.** The air-magnet single coil structure.

The coils are transported with some simple rigid tool and assembled by positioning one on top of the other and connecting them by using screws. Periodically a check of the vertical flatness must be performed and eventually shimming must be done to avoid deviation from the vertical line (Figure 44).

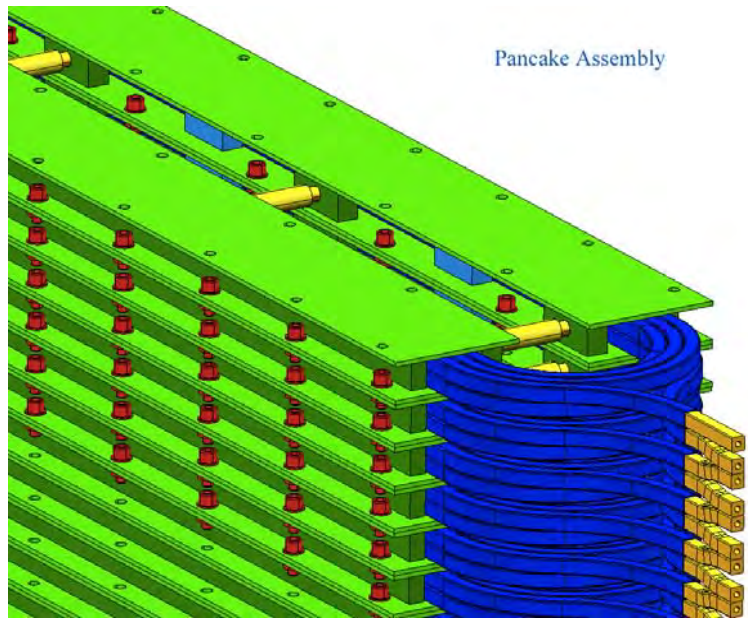

**Figure 44.** The air-magnet coils assembly.

### 3.2.2 *Analysis method*

The magnet design was assisted by Finite Elements (FE) calculations. Preliminary results were obtained in 2-D computations using the COMSOL program. These calculations were followed by full 3-D analysis, mainly with the Vector Fields OPERA program. As previously said a much more detailed magnetic and structural analysis will be done to reach and validate the final design.

#### *Field Profiles*

We choose a right-handed co-ordinate system with the origin at the interaction point and the z-axis along the beam direction, pointing towards the muon system; the y-axis points upward and x horizontally. The symmetry axes of the dipole follow the same directions, with the principal field component lying along y. The line connecting the centers of the pole faces passes through z = 5.3 m. The field is oriented along the y direction. Its uniformity was evaluated along straight tracks originating at the interaction point. The variations relative to the central track are below 0.1%. The fringe field is much dependent on the specific geometry of the iron and the coil structures. A careful optimization of them will allow to obtain the fringe field values below the constraints imposed by the LAr electronics. Figure 45 depicts the magnetic field in a 3-D model, while the transverse size of the global magnetic field (in air and in iron) is shown in Figure 45a.

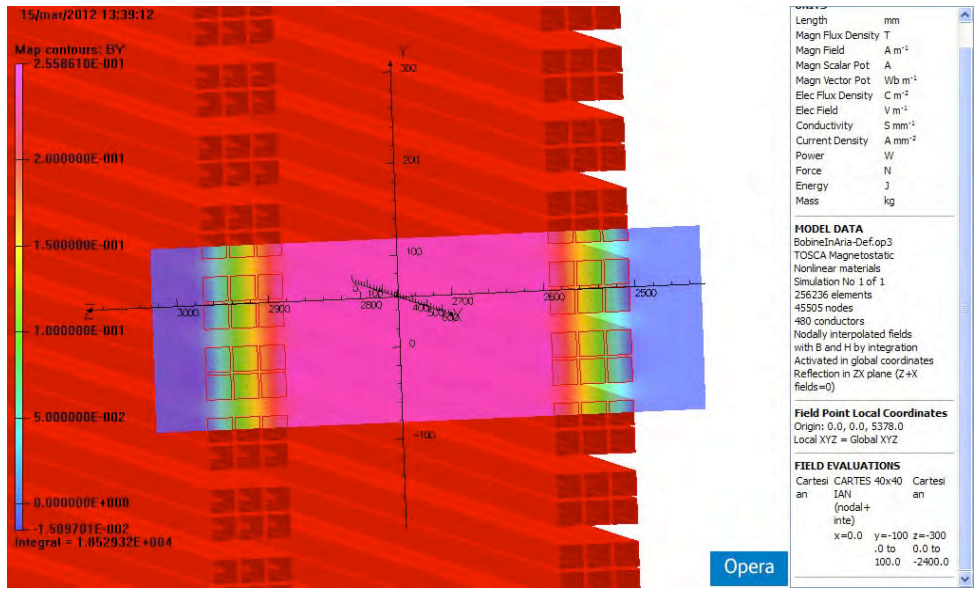

**Figure 45.** The air-magnet field map in a 3-D model.

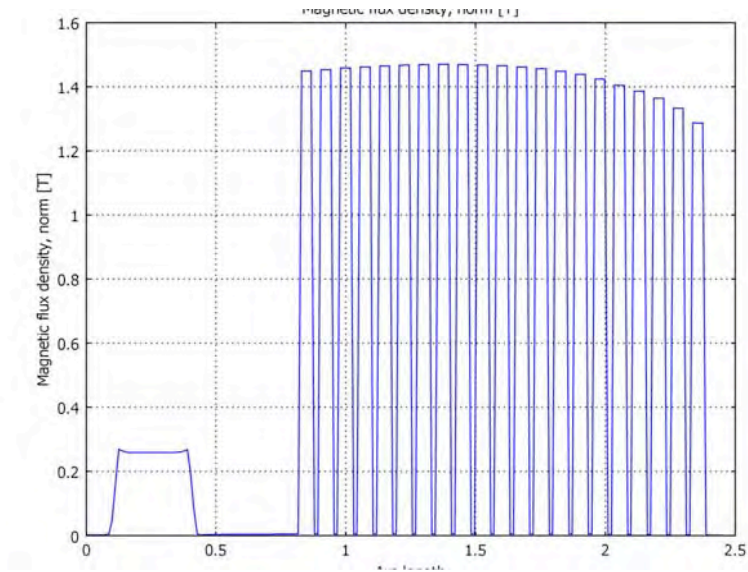

**Figure 45a.** The transverse profile of the global (air plus iron) magnetic field.

### 3.3    *The Detectors*

#### 3.3.1    *Detectors for the iron magnets.*

The NESSiE Near and Far spectrometers will be instrumented with large area detectors, for precision tracking of muon paths, high momentum resolution and charge identification capabilities.

Suitable active detectors for the iron magnets are Resistive Plate Chambers (RPCs) with plastic laminate electrodes, a well established technology for building large area detectors at an affordable price, presently in use at LHC ([17], [18] and [19]) as well as in neutrino and astroparticle physics experiments ([20], [21]). RPCs are gaseous ionization detectors with parallel resistive electrode plates, as sketched in Figure 46. A single gas-filled gap delimited by bakelite resistive electrodes is the simplest set-up commonly used in streamer mode and digital read-out. To ensure a uniform electric field, a lattice of spacers is placed inside the gas gap with 10 cm pitch.

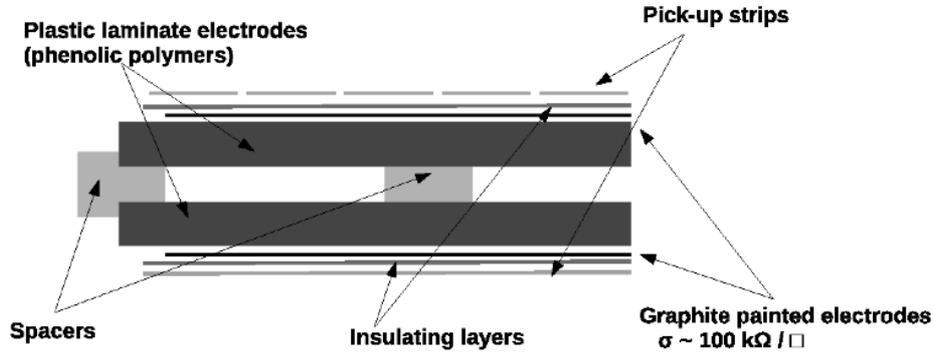

**Figure 46.** Resistive Plate Chamber sketch.

The detector design for the NESSiE magnets is based on the RPCs developed for the OPERA experiment at LNGS. Each chamber has dimensions of 2904 x 1128 mm$^2$. In the Near spectrometer the internal RPCs will be arranged in planes of 2 columns and 3 rows, for a total exposed surface of about 20 m$^2$. A total of 40 layers will be instrumented with 240 chambers. In the Far spectrometer the internal RPCs will be arranged in planes of 3 columns and 5 rows, to form planes of about 50 m$^2$. A total of 40 layers will be instrumented with 600 chambers.

Given the expected counting rate, comparable to that from cosmic rays, plates with resistivity between $10^{11}$ and $5 \times 10^{12}$ Ohm*cm at T=20ºC are suited and the streamer operation regime, with high amplitude signals, can be used. A space resolution of about 1 cm, sufficient for the detectors inside the iron magnets, can be achieved by using 2-3 cm pitch copper strips digitally read out. Their orientation and size are optimized to reduce ambiguities in hit reconstruction (especially in the Near spectrometer).

The OPERA RPC system, made of about 1000 chambers for a total surface of about 3000 m$^2$, is operated with a gas mixture composed of $Ar/C_2H_2F_4/I$-$C_4H_{10}/SF_6$ in the volume ratios 75.4/20.0/4.0/0.6 [22]. The detectors are flushed at five refills/day with an open flow system. With this gas mixture the RPC are operated at the voltage of 5.8 kV with a current of less than 100 nA/m$^2$. The high voltage (HV) is symmetrically split between the positive and the negative electrode to ease insulation from ground.

Typical tracking resolutions of OPERA RPCs are shown in Figure 47. A more general description of detector performances are reported in [23], [24] and [25]. For the gas system at CERN we could profit of the experience on recirculating gas systems to increase the gas flow rate and reduce gas consumption.

About 1000 m$^2$ of RPCs are available at Gran Sasso as unused remainders of the OPERA production. Plastic laminates for a new production can be produced in Italy by Pulicelli company, located near Pavia and currently produc-ing material for the CMS RPC upgrade. RPC chambers can be assembled in Italy as well, at the renovated General Tecnica company. The Quality Control test setup, used for OPERA detectors [26] is still partially available at the Gran Sasso Lab.

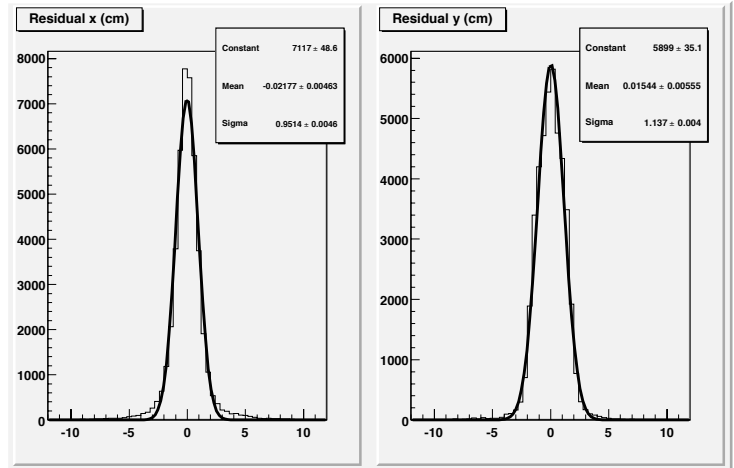

**Figure 47.** Tracking resolution for the bending (left plot)  and for the othogonal coordinates (right plot).

The front-end electronics (discriminators based on LVDS receiver circuits) and the high voltage distribution (including the dedicated nanoamperometers designed by the Electronics Workshop of LNF [26]) can be recovered from the OPERA experiment. The OPERA DAQ is a free-streaming system; for NESSiE we plan to have a trigger based on the proton extraction early warning. A spectrometer trigger based on a majority of the RPC layer OR signal can also be implemented for monitoring the detector performances on cosmic rays.

### 3.3.2 *Detectors for the Air magnets.*

The air-core magnet will be used for the charge identification of low momentum muons which requires precise measurements of the muon path. Track reconstruction is required only in the bending plane (x-z view). The identification of the muon charge is optimized by looking at the change of the track slope before and after the magnetic field. A spatial resolution of the order of 1 mm is adequate, as reported in [3].

Different detector options are available for such precise measurements. In the following the choice of Resistive Plate Chambers with analog read-out will be discussed. Such detectors reach the required performances at a limited cost. Finally we would like to stress that the RPCs are used also in the iron section although in digital read-out. Therefore this choice allows to have only one kind of detector in the whole spectrometer.

In addition to the analog read-out on the x-z view, the digital read-out will be implemented also on the y-z view with a restrained cost. This additional view will allow to reconstruct 3D tracks and to reject spurious hits due to electronic noise or to cosmic rays.

### 3.3.3 RPCs with analog read-out

As previously reported RPCs are gas detectors [27] widely used in high energy and astroparticle experiments. The most relevant features are the excellent time resolution (ns) and the high rate capability. Also the position resolution is very good, in particular conditions the centroid of the induced charge profile was determined [28] with a FWHM resolution of $\sim 0.12$ mm. In the case of the NESSiE experiment such resolution is not required and a simpler and cheaper set-up can be adopted.

The analog read-out of RPC has been implemented in the last years [29] or variously proposed[2]. With this technique, by reading the total amount of charge induced on the strips, detailed information can be obtained on the streamer charge distribution across the strips and better estimate of the track across the detector is thus achieved than in the digital case. The charge profile can be approximated by a Gaussian shape whose width ($\sim 5$ mm) does not depend on gas mixture and operating high voltage, unlike the total charge which is strongly dependent on them (see Figure 48).

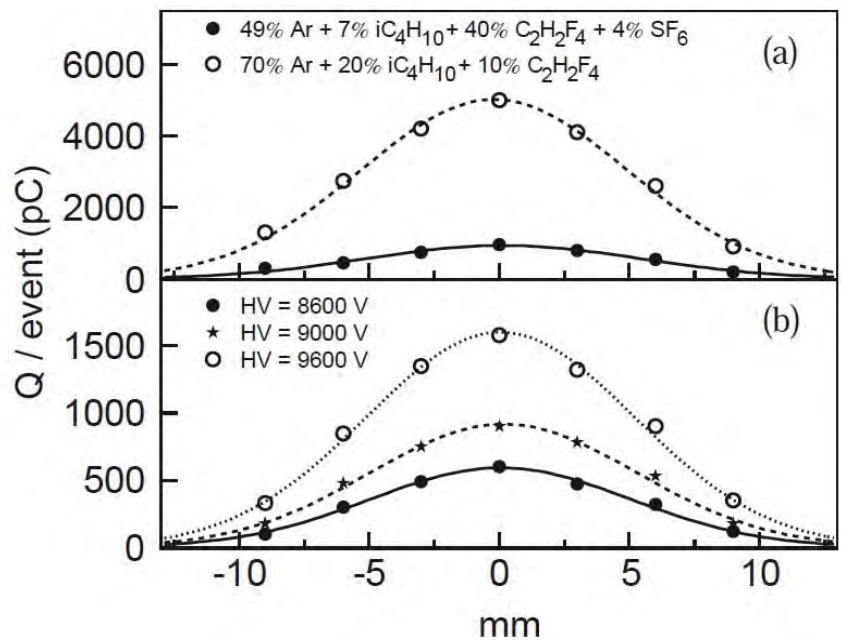

**Figure 48.** Charge profile (normalized per event) observed with RPC working in streamer mode (analog read-out and 2-mm strips): (a) for various gas mixtures at nominal running HV, (b) for different high voltages with the gas mixture (49%Ar + 7%iC4H10 + 40%C2H2F4 + 4%SF6) foreseen for the ALICE detector. The curves are Gaussian fits of the data. The standard deviation of these distributions is about 5 mm in all cases [28].

A few mm resolution in the charge position determination is obtained by choosing an adequate strip size (of about 1 cm). Also the dynamic range is improved, allowing the detection of particles at a density of the order of 1000 particles/$m^2$.

The detection efficiency is crucial for this measurement, it must be higher than 90% and large dead zones are not allowed. In order to increase the efficiency of each RPC plane the set-up shown in Figure 49 has been chosen. Cables

---

[2] The analog read-out of RPCs strips was for example proposed as an alternative option for the Target Trackers of the OPERA experiment [27]

and connections for HV, gas and front-end electronics are reduced as much as possible according to the OPERA set-up [32].

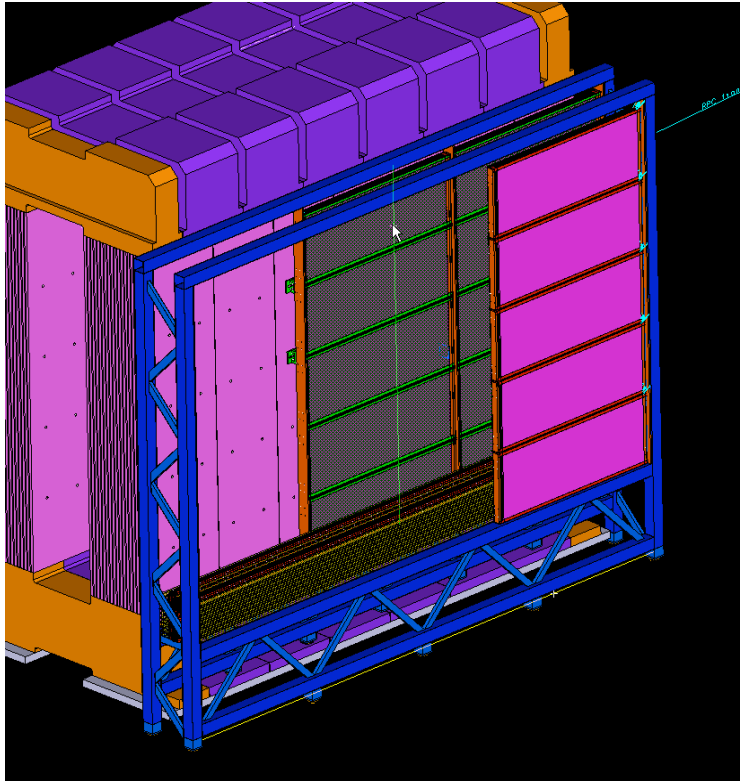

**Figure 49.** Set-up of RPC planes for the air-magnet.

For the low momentum particle charge measurement 4 RPC planes upstream and 4 RPC planes downstream the magnetic field in air, respectively, are required. The single RPC surface is $2.91 \times 1.13$ m$^2$, each plane in the Near (Far) Detector is arranged in 3 rows and 2 columns (5 rows and 3 columns) for a total of 48 (120) RPCs. The RPC are operated in streamer mode with the same gas mixture as for the RPCs inside the Iron-Magnet.

### 3.3.4 The Electronics

A digital read-out is foreseen for the RPCs instrumenting the iron magnets.

According to the same design scheme adopted for the OPERA experiment [29], signals coming from RPCs working in streamer mode will be directly read-out by means of LVDS receivers, without the need of preamplifiers. Since RPCs will be embedded in the iron structure, the front-end boards (FEB) collecting signals from the strips will be housed in a counting room located on the top of the magnet, as done in OPERA, in order to ease the system maintenance. The receivers will act also as discriminators with programmable thresholds.

The expected rate in the spectrometer will be of several tens of events *per spill* in the Near detector. Due to the high event rate, the FEBs will be operated in trigger-less mode. The discriminator output will be sampled and stored in a circular buffer driven by an external clock for the duration of an external spill gate, meant to include the beam spill. The spill gate will be generated by the *start-of-spill* and the *end-of-spill* signals, both time-stamped to uniquely identify the event, the latter triggering the data acquisition process. Data will be transmitted to the DAQ during the inter-spill time, via an Ethernet interface integrated on each FEB.

The same operation mode will be implemented in the Far detector, where only a few events per extraction are expected.

About 32,000 channels in total (Near and Far spectrometers) will be used to read-out 2,500 $m^2$ of RPC detectors with pickup strips of 2.6 cm pitch in the vertical direction (orthogonal to the bending plane) and 3.5 cm pitch in the horizontal direction (tracking without bending).

The strip read-out will be organized per plane. About 340/500 channels per plane are considered for the near and the far detectors, respectively.

Each FEB board will house 64 channels and will produce a FAST-OR trigger signal if at least one of the read-out strips is fired. A Trigger Board (TB) will collect the FAST-OR outputs from FEBs and will generate a programmable trigger which can be used for the acquisition of cosmic ray muons and for monitoring and calibration purposes.

A total of 600 FEBs will be produced. They will be based on the use of FPGAs (Field Programmable Gate Arrays) implementing the circular buffer for the event storage, the time-stamping, the FAST-OR with noisy channels masking and the Ethernet interface. FEB parameters (thresholds, masks, etc) will be set through the Ethernet interface.

Assuming 64 channels per board, a clock of the order of 10 ns and a spill duration of the order of 10 $\mu$sec, the event data per FEB will be of the order of 10 KB.

The front-end and trigger boards will be designed by electronics CAD service.

Among the possible choices for the air-core magnet detectors, the use of RPCs working in streamer mode with analog read-out is first envisaged. The development of different front-end boards based on ASIC chips with 10-bit ADCs is being considered.

The ASIC chip will house 64 channels. Each channel will consist of an electronic chain with a preamplifier, shaping the RPC signals at about 200 ns, a flash ADC with a sampling rate of 25 ns, and a memory allowing storing time-stamped data after zero suppression. An Ethernet interface will be used for the communication with the DAQ. A trigger-less mode of operation will be implemented, as well as a triggered mode (with the digital FAST-OR), as discussed above for the FEBs to be developed for the RPCs with digital read-out.

# 4   Data taking

### 4.1    DAQ for LAr and Trigger system.

The event building architecture of the T600 is based on a network characterized by a two level switching layers (Figure 50). CPU's on the readout crates are connected through a 100 Mbps Fast/Ethernet link to a first 24 ports switch. A central 1 Gbps Giga/Ethernet switch then connects all the local switches to the PC farm.

The layout provides a receiving and merging workstation handling each readout chamber (24 readout units), with a maximum input throughput of ~ 50 MB/s and safely exploiting the link at half of the available bandwidth. All the readout units can work autonomously, pushing their own data to the receiving workstation. Segmentation and parallelization of the data stream (e.g. 12 readout units per builder unit) allow reaching a building rate > 1 Hz on the whole T600, largely adequate to match the SPS extraction rate.

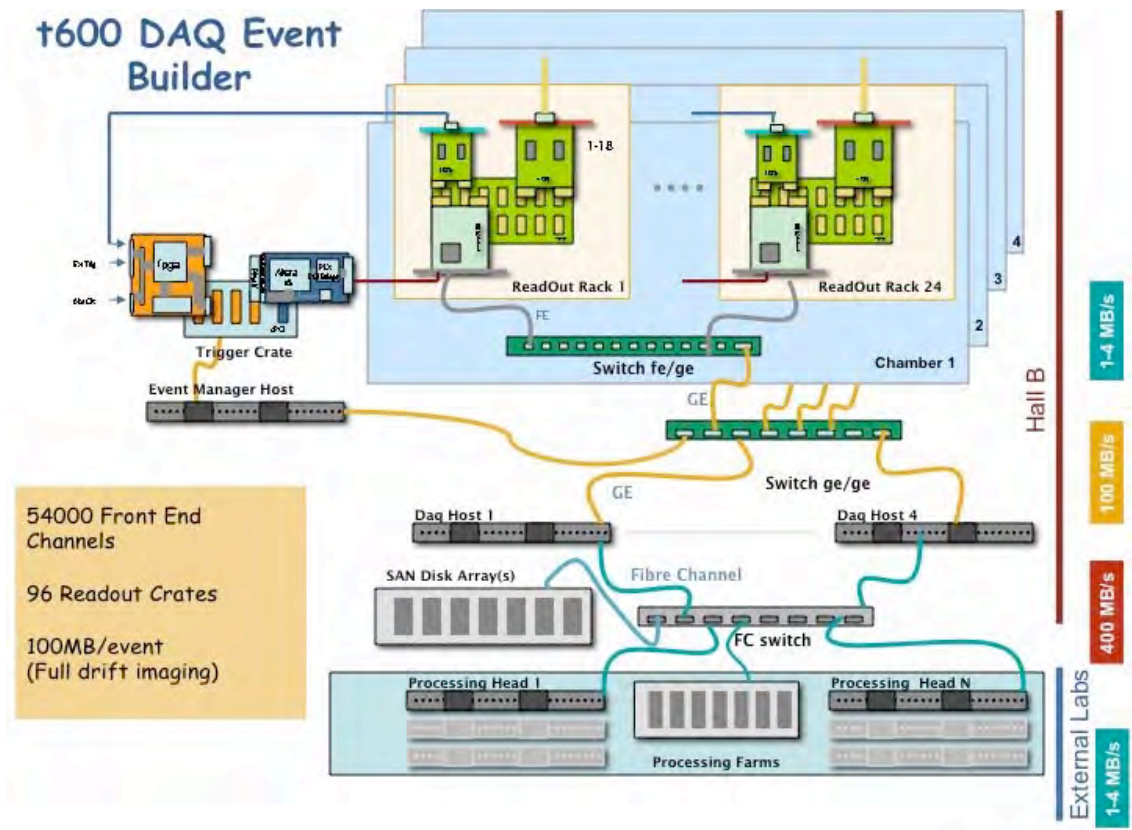

**Figure 50.** The T600 Data Acquisition system architecture.

The same DAQ architecture will be implemented in the T150 detector at the Near site. The trigger system will exploit the spill gate in coincidence with the signals of the PMT of each TPC.

At the nominal beam power, ~ 0.18 (0.1) neutrino interactions per spill with vertex in a LAr-TPC's are expected in the Far position for positive (negative) focusing, with a slightly lower rate of beam related through going muons. Therefore the coincidence of the PMT signals within the spill will be used to trigger events acquisition in the T600.

Because of the expected high rate of neutrino interactions in the T150 detector (more then 1 event/spill), the beam gate alone will be used to trigger the data collection.

According to the average SPS repetition rate (1 spill/6 s), ~ 1.25 $10^4$ trigger/day on the near detector and the far detector are expected. The corresponding data volume is around 0.5 TB/day, recording only the TPC where the event is localized and assuming the present data compression algorithms. This is comparable with the present T600 data-taking throughput.

### 4.2    DAQ for RPC detectors

The Data Acquisition system is built like an Ethernet network whose nodes are the FEB equipped with an Ethernet controller. The Ethernet network is used to collect the data from the FEB, send them to the event building workstation and dispatch the commands to the FEBs for configuration, monitoring and slow control.

This scheme implies the distribution of a global clock to synchronize the local counters running on the FEBs that are used to time stamp the data. The clock signal can be distributed either using a dedicated network or with the same Ethernet network implementing a White Rabbit protocol [33].

The DAQ clock is synchronized with the CERN General Time Machine signal in order to start the DAQ readout cycle during the proton extraction time.

As described in the Section 3.3.4, along the spill duration the FEBs store the status of the discriminators in a circular buffer driven by an external clock. The readout of the buffer by the DAQ is triggered by the *end-of-spill* signal sent to each FEB. This signal causes the FEB to disable writes and the buffer content is transferred to the Ethernet controller. Here data are time stamped, eventually zero suppressed and sent through Ethernet to the event building. The *start-of-spill* signal is used to abort all the readout process on the FEBs and to start new data read-out.

In the inter-spill time the acquisition of cosmic ray muons and calibration data is triggered by a fake spill gate, possibly validated by a programmable logic (Trigger Board) on the basis of the FAST-OR signals generated by the FEBs.

Assuming 64 channels per board, a clock of 10 ns and a buffer depth of 2000, needed to acquire 20 μsec of data, the time needed to transfer the data to the event building is less than 200 μsec on a Gigabit Ethernet.

### 4.3    LAr data reduction.

Based on the experience of the ICARUS Experiment at LNGS, data handling (quality monitoring, filtering, prompt reconstruction and event tagging) and data streaming (DAQ to central storage facility) will duplicate the present T600 scheme based on local live buffering disk space. As far as data storage and computing power resources are concerned the use of the already existing CERN facilities (for instance CASTOR and GRID) will be exploited.

To quantify the amount of disk space and CPU power needed by the experiment, a total rate of about $2.5 \cdot 10^6$ triggers/year is assumed both in the Near and Far detectors. On the other hand, the size of an event after online lossless compression is of about 50 Mbytes in the Far detector and ¼ of that in the Near one. This translates to a total of about 160 TBytes of disk space needed for each year of data taking:

$2.5 \cdot 10^6$ triggers/year * (12.5 MB $_{Near}$ + 50 MB $_{Far}$) = 156 TB/year.

In case only one module of the Far detector is readout per each trigger, the occupancy is reduced to ~ 100 TBytes per year. As a comparison, the total data rate of the ICARUS Experiment at LNGS is already ~ 100 TBytes per year.

The amount of CPU time needed in order to process one full drift T600 event in the ICARUS experiment is of about 2 minutes using a machine having a value of 40 HEP-SPEC06 [34] units (HS06); this includes filtering and hit finding. The average processing for the Near detector events will then require 30 s. For a total of $2.5 \cdot 10^6$ events, both in Near and Far detectors. The expected amount of CPU power to process one full year of data in three months is of about 50 machines with 40 HEP-SPEC06 units (2000 HS06):

- Processing time: (120 s $_{Far}$ + 30 s $_{Near}$)/event * $2.5 \cdot 10^6$ events = $3.8 \cdot 10^8$ s;

- $3.8 \cdot 10^8$ s / 50 cpu$_{40\text{-}HS06}$ ~ 3 months.

As a comparison, the ICARUS processing farm at LNGS is capable of 240 HS06 units. These figures of merit in terms of disk occupancy and CPU power could be further improved taking into account that in both the Near and Far detectors the DAQ system will benefit of the Region of Interest (ROI) search performed on-line with the Super-Daedalus chip which allows to reduce considerably the quantity of data per event [35].

# 5   Installation at CERN

In the SPS beam layout, two new halls have to be built along the neutrino beam line within the CERN north area (see Figure 1), hosting the T600 and the T150 with the corresponding Spectrometers. In such halls a possible solution to match the beam level could consist in excavating pits of adequate size, where both the LAr-TPC's and the Spectrometers will be located.

These new halls must be equipped with electrical power, air ventilation, water cooling and crane to lift small weight objects. The total electrical power required to operate the T600 detector is about 600 kW (including 100 kW spare power) while for the T150 about 350 kW will be needed (including 50 kW spare). The total electrical power required to operate the spectrometer magnets is about 3 MW for the Far site and 2 MW for the Near site.

## 5.1   *Far site.*

The general layout proposed for the Far site experimental hall is given in Figure 51. The LAr-TPC is placed upstream with respect to the neutrino beam direction. The air magnet part of the NESSiE spectrometer is located immediately after. The air magnet allows to track muons emerging with low momentum from the LAr volume. Further downstream, the iron magnet, designed to track higher momentum muons, is installed.

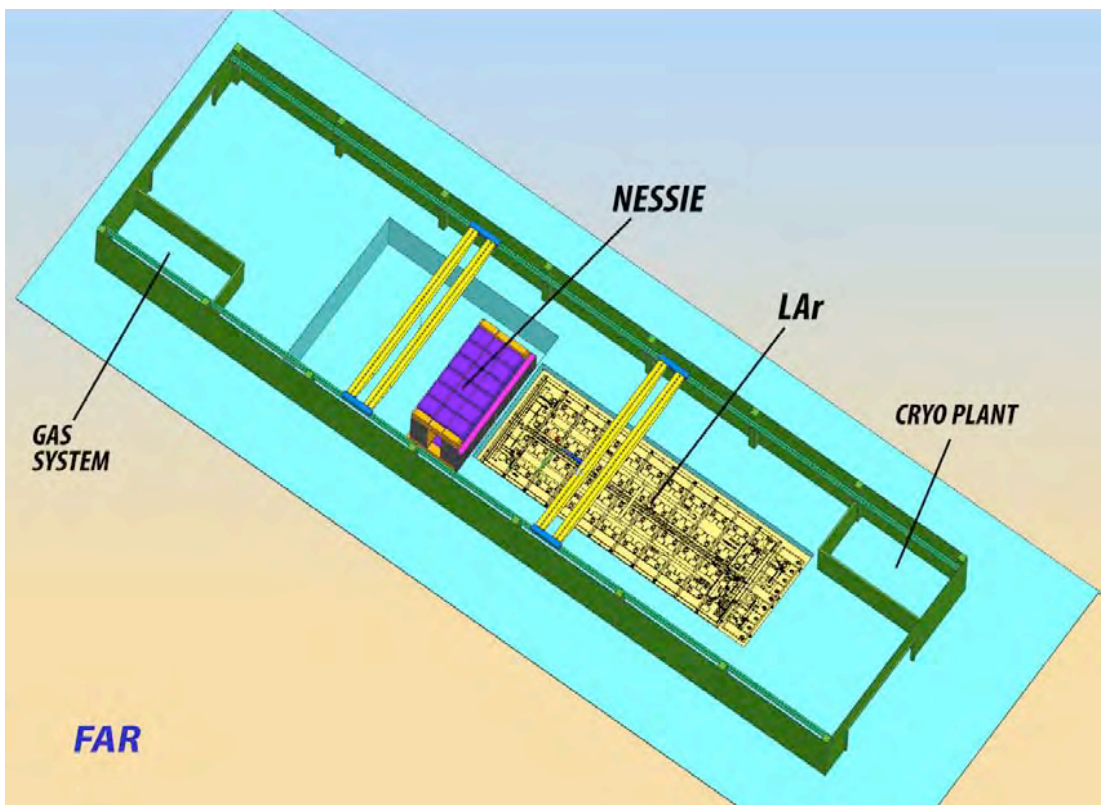

**Figure 51.** General layout of the Far site experimental hall.

The LAr and the Spectrometer detectors are placed in a 8 m deep pit, whose concrete walls could be used to hold the insulation layers of the LAr-TPC following the technique depicted in the Section 2. The LAr detector vessel and the thermal shield could be inserted from the top. An additional floor above ground is envisaged to host the detector external equipment (electronics, cryo-coolers nitrogen storages). The LAr-TPC and the Spectrometer are separated by a containment concrete wall. Additional ancillary systems are foreseen: two cranes needed for detectors assembly, LAr cryogenic plant and Spectrometer gas system. For the final positioning of the LAr detector two 30 t cranes will be simultaneously needed. For the iron magnet assembly one of the two 30 t cranes will be used continuously. A possible additional 5 t crane to be used only part-time (not shown in Figure 51) will help in speeding up the assembly.

More details are shown in Figure 52. The volume required for the installation of the T600 LAr-TPC modules is 24 m (length) x 12 m (width) x 7 m (height). A volume of 15 m (length) x 13 m (wide) x 8 m (deep) is foreseen for the assembly of the air and iron magnets of the NESSiE spectrometers. The iron magnet will be more easily built at about 5 m from its nominal position along the beam axis and then translated to its final position together with the air magnet. An additional volume of 15 m (length) x 17 m (wide) x 8 m (height) is foreseen for the initial storage of the iron slabs and other ancillary systems. Later on this area will be occupied by the gas system and by the PC farm for the data acquisition system.

The slightly different depth of the LAr-TPC and Spectrometer pits is mandatory to have the detectors horizontally centered. In Figure 52 the support and thermal insulation structures of the LAr-TPC detector are not shown. Keeping also in mind the space constraints of the LAr the overall volume of the Far experimental hall will be 64 m (length) x 17 m (width) x 8 m (height).

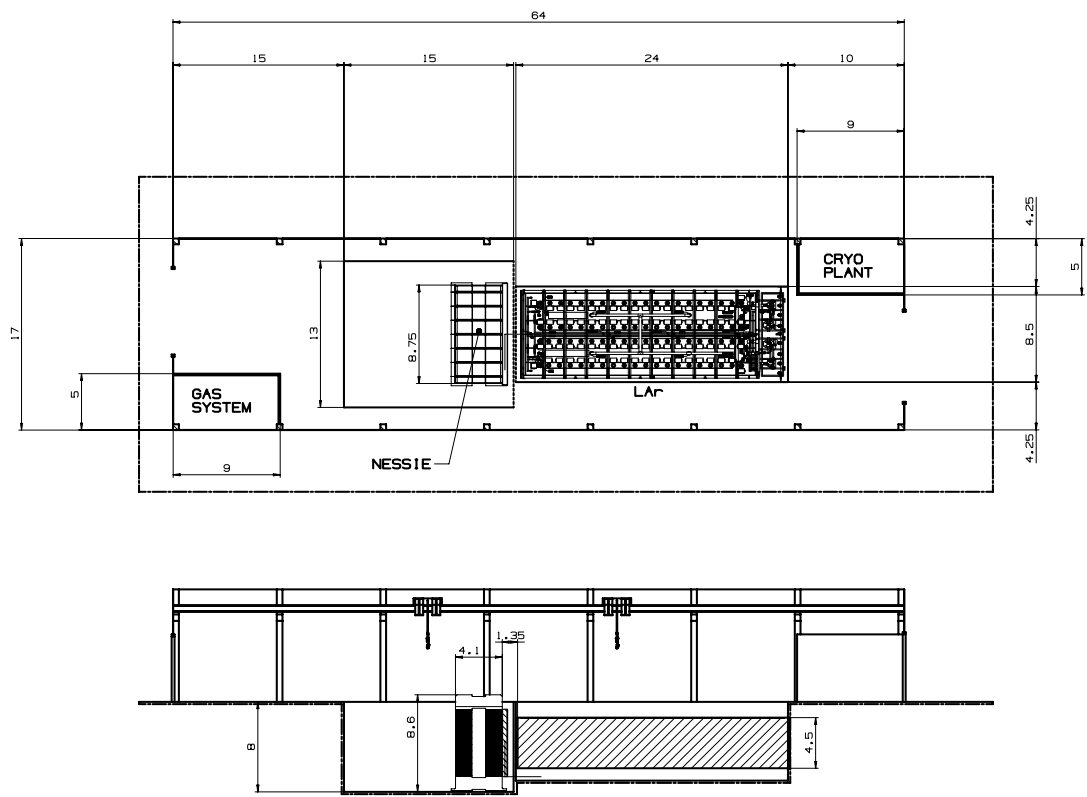

**Figure 52.** Preliminary details about the occupancy of the Far experimental hall. At present the details of the LAr detector are available only for the top view.

### 5.2    Near site

The layout of Near site hall is obviously similar to that of the Far one (Figure 53). The pit volume required for the installation of the T150 LAr-TPC module is 14 m (length) x 8 m (width) x 7 m (height). The thickness of the spectrometers at the Far and Near sites is the same, hence also the length of the pit is 15 m. Due to the reduced transverse dimensions of the Near iron magnet the pit is only 9 m wide. Downstream from the pit area a volume of 10 m (length) x 17 m (width) x 7 m (height) is foreseen for the initial storage of the iron slabs and other ancillary systems. Later on this area will be occupied by the gas system and other support structures for the data acquisition (PC farm).

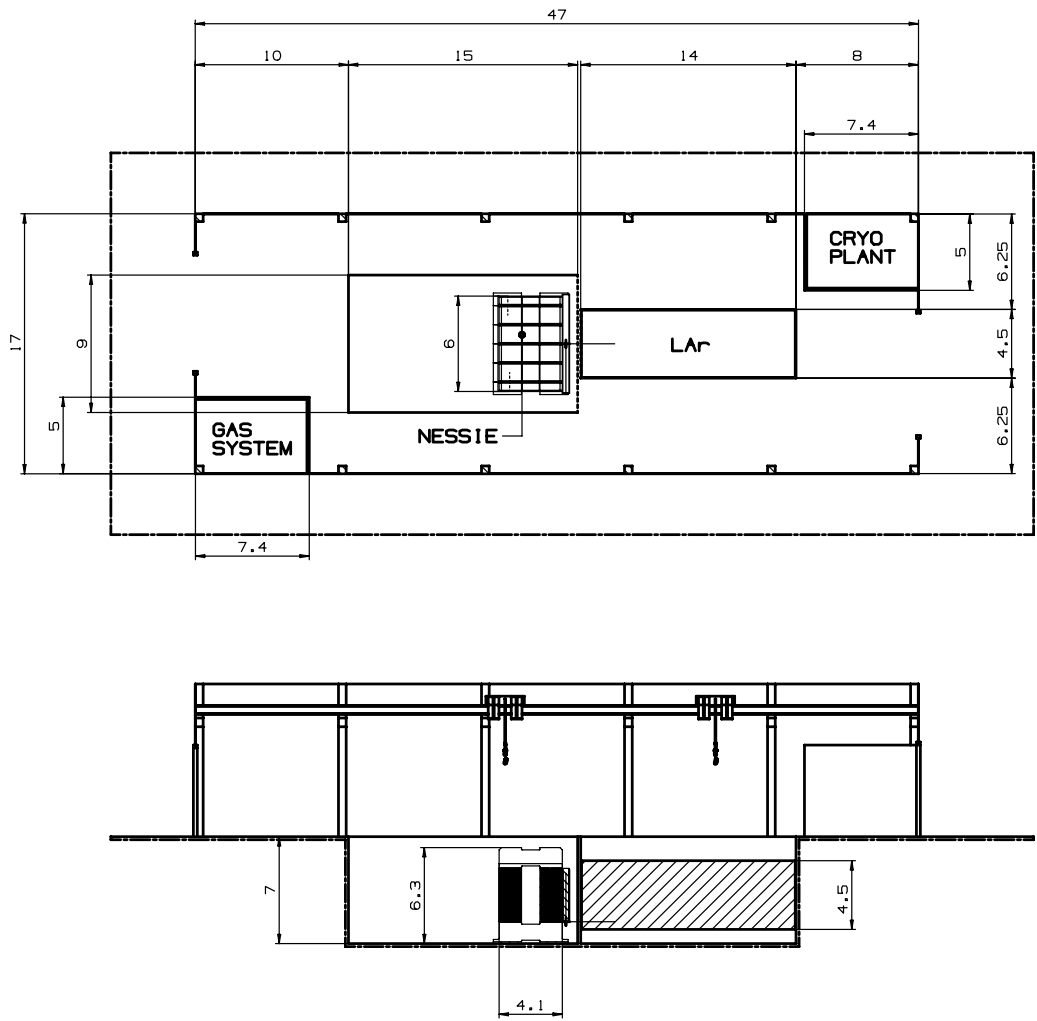

**Figure 53.** Preliminary details about the occupancy of the Near experimental hall.

### 5.3    Logistics

An area of approximate dimensions 45 (l) x 7 (w) x 6 (h) m$^3$, will be required to perform all the assembly operations for the two T600 modules and for the T150. Half of the area will be equipped with a clean room (10000 class) facing the aluminium vessels. In the clean room will take place all the operations required for the inspections and upgrades of the T600 internal detector and the assembly of the T150 internal detector before sliding them inside the aluminium vessels. We note that the characteristics of Hall 191 match the requirements for the assembly area also in terms of available equipment.

### 5.4    Assembly procedures

As mentioned in the previous paragraph assembly of internal detector components and insertion in the aluminium vessels will take place in a dedicated area equipped with a clean room of adequate size. When this area will be ready and the first vessel will be available for delivery to CERN, transport of the first T600 internal detector will be performed. The internal detector will be placed in the new clean room where it will be prepared for its insertion in the new vessel. After the positioning and completion of the internal detector assembly, the vessel will be disconnected from the clean room and the front door will be closed and sealed. Part of the internal cabling will also be performed (installation of the chimneys) before moving the detector to the final location. A helium leak test then will be done to check all the connections that will become inaccessible (the front opening, the chimneys and the connections for LAr recirculation and emptying). Operations will be organized in the same way for the second T600 module. Assembly and insertion of the T150 internal detector will be done at the end.

Preparation of the new experimental sites will proceed in parallel. As soon as the civil works will be completed, starting from the far detector site, assembly of the external insulation will begin. Bottom and lateral sides will be assembled. Bottom and lateral sides of the thermal shield will be installed. The vessels will then be inserted from the top using two cranes.

The top part of the thermal shield will be then installed and the complete shield will be tested for tightness. The top part of the insulation will be finally assembled and the top floor for electronics and other equipment positioning will be put in place.

The internal cabling will be completed with the installation of the feed-throughs. The argon and nitrogen circuits will be connected and completed. All the circuits directly connected to the principal argon vessels will be tested at room temperature while the Nitrogen re-condensers will be operated in a closed loop with the main nitrogen storage tank. The electronic racks will be put in place, but not connected to the feed-throughs; the final connection will be performed after the cool down.

At this point, the commissioning will start with the vacuum phase. The total time needed to complete the commissioning up to the detector filling will be about three months.

As a tentative schedule, in Figures 54 and 55 the operation times for the LAr-TPC and the Spectrometers are respectively shown.

### 5.5    Safety issues

Considering the severe constraints imposed by the handling and storage in an underground experimental cavern of a huge quantity (770 tons) of lique-fied Argon, the safety issues have been the subject of special attention during the design and construction of ICARUS.

The Maximum Credible Accident (MCA) that has been considered for the T600 operation in LNGS is the possibility of a significant mechanical failure of the principal argon vessels with consequent huge spillage of LAr. This lead to the construction of an external cryogenic containment vessel (the thermal insulation) and a dedicated gas exhaust system. The design of the plant was done with the anti-seismic prescription of the Italian regulation for the LNGS area, that has seismic degree 9, the highest in Italy. The whole T600 assembly is supported with special anti-seismic dumpers (shock absorbers) and all transfer lines are equipped with flexible joints to allow for reciprocal movements.

Furthermore, safety reviews, a detailed risk analysis and safety audits have been carried out by external experts in collaboration with the LNGS staff prior and during the installation phase in the underground hall B of LNGS.

As a result of the design studies and of the outcome of the reviews, to mitigate the consequences of a possible failure the following protections and safety procedures have been implemented:

- Cryogenic containment vessel to prevent large liquid spillages;
- Anti-seismic special feet;
- Safety disks and overpressure valves;
- Dedicated gas extraction system with heaters for cold gas exhaust;
- Extensive installed redundancies (Control PLC, pumps, valves, etc.);
- Video remote monitoring, local and remote alarm system and audio warnings;
- Individual protections, individual and fixed oxygen deficiency measurements, dedicated escape ways and ventilation system.

Procedures are as follows:

- Training of personnel;
- Evacuation plan;
- Minimisation of the presence of personnel in the experimental hall B;
- Inspection and maintenance.

Considering that an underground laboratory like the LNGS is a hostile environment for such a huge cryogenic installation, we believe that the design features and the safety precautions that have been implemented for ICARUS at LNGS will also meet, after their adaptation to the environmental conditions of the CERN Laboratory, the CERN safety requirements.

Similar arguments can be evoked for the seismic issues. We estimate that the protections adopted at the LNGS will fully satisfy the anti-seismic rules applied by the CERN Laboratory.

We stress in particular that the basic design of the ICARUS cryostat, with two separated internal liquid argon containers surrounded by a common skin kept at the same temperature as the bulk of the liquid and acting as containment vessel, is maintained. This will be achieved by the adoption of the membrane tanks design described in 3.4.3. Moreover the membrane tank solution described in this proposal offers an even higher safety level than what adopted at LNGS since it is fully passive, i.e. it does not require the vacuum in order to achieve an efficient thermal insulation.

Finally, all the new vessels will be designed and tested according to the European Pressure Equipment Directives (PED). Moreover those vessels that will be recovered (LN$_2$ back-up tanks, nitrogen shield, LN$_2$ phase separators and condensers, purifiers etc.) will be pressure tested again after transportation and installation at CERN according to their design pressure.

### 5.5.1  Spectrometer Magnets

The general safety is based on the Magnet Safety System (MSS), as part of the Experiment Control System (ECS).

The magnet power supply is interlocked to the general fire alarm and to the specific water leak alarm for the magnet. The boxes used for protection against electrical hazards also serve as protection against water spills.

These protection boxes are equipped with water leak sensors and are connected to a sink.

#### Mechanical Safety

All the construction parts, supports and tools required for the assembly will follow European directives, CERN safety codes and rules and the CERN safety policy expressed in the SAPOCO document [36]. General criteria for safe stress levels will follow European and/or international structural engineering codes stated in Eurocode3 [37] and AISC [38].

For anchoring points, lifting and rigging gears, which will remain the property of CERN/NESSiE, the design, manufacture and testing shall comply with the CERN Safety Code on Lifting Equipment (Safety Code D1 [36]) and construction norms and codes

#### Electrical safety

The conductor insulation will conform to CERN Safety Instructions IS23 and IS41 [36] and will be tested to 3 kV. The power supply will be of the insulated IT-type (IEC364 and IS24 [36]). Low voltage safety rules will apply according to CERN Safety Instructions IS33 [36] and IS24 [36].

As protection against electrical hazards, the region of the coils with the electrical interconnections of the pancakes will be completely surrounded by a polycarbonate box. This includes also the flexible water connections.

#### Safety in the magnetic field

Safety requirements stipulated in CERN Safety Instruction IS36 [36] will be respected. The general safety system will be based on the Magnet Safety System (MSS) and the Experiment Control System (ECS/DCS).

# 6 Time Schedule

The two scheduled projects, separated for the LAr-TPC and the Spectrometer parts, are shown in Figures 54 and 55, respectively. The foreseen times needed for the T600 LAr-TPC transportation from LNGS and arrangements at CERN, as well as the construction and assembling of the two spectrometers, are fitted into a 2 years window, as soon as the starting will be provided by CERN and Funding Agencies.

More refined planning and cost estimates will be finalized once the corresponding Scientific Boards and Funding Agencies will approve the project and we will then receive "green light".

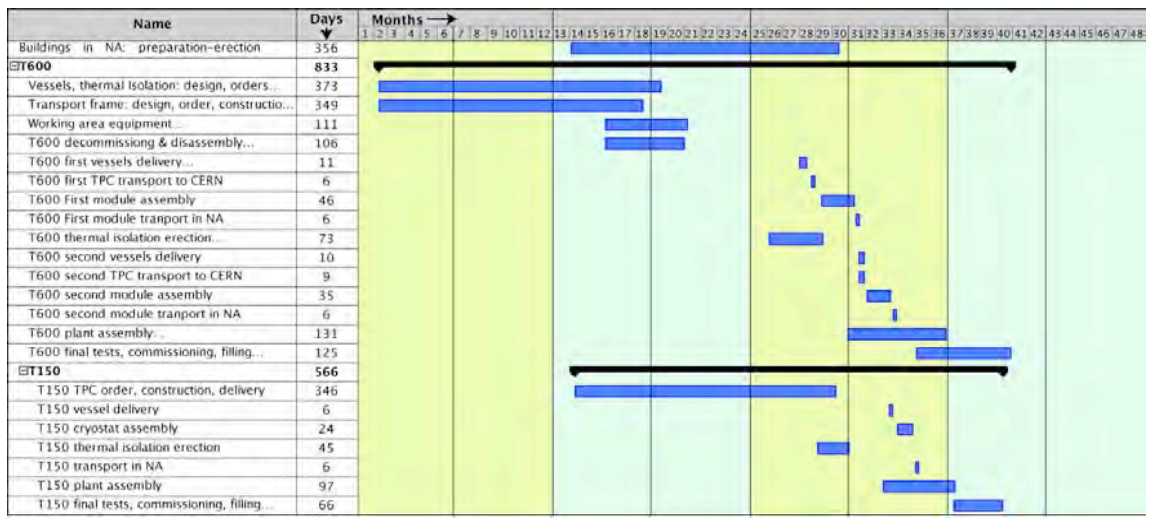

**Figure 54.** Schedule Project, LAr part.

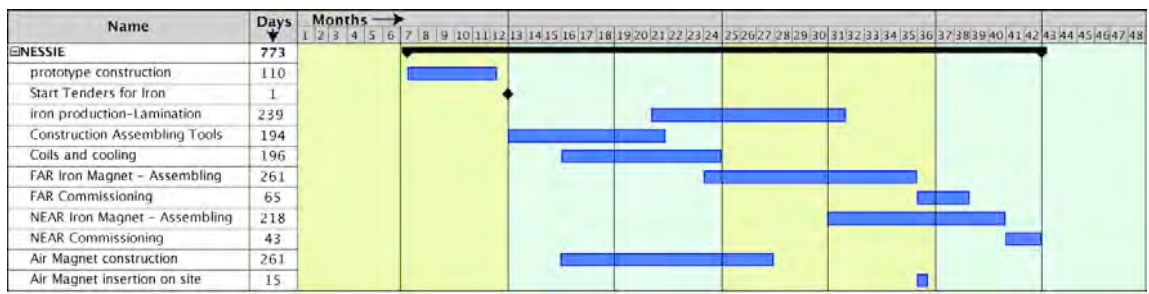

**Figure 55.** Schedule Project, Spectrometer part.

# 7 Acknowledgements

We acknowledge the valuable support of, and discussions with, the CERN Management and different Departments. The Collaborations are especially grateful to the Director General, the Directors for Research and Accelerators and the SPS-C Board for their encouragement; the BE, TE, EN, GS and DGS Departments gave a technical effective contribution in guiding through delicate choices. In particular we whish to thank C. Bertone, M. Calviani, I. Efthymiopoulos, J.L. Grenard, R. Steerenber, R. Trant.

We express our appreciation for the support and encouragement of INFN and especially of LNGS Director, Commissione II, its President and our appointed referees.

J.M. Didier (LNGS) provided important contributions to the cryogenic plant project; A. Viale (Genova) gave valuable support in defining the air magnet.

# 8 References

[1] C. Rubbia et al., ICARUS/CNGS2 Collaboration, *"Physics Programme for ICARUS after 2012"*, SPSC-M-773 (2011).

[2] C. Rubbia et al., *"A comprehensive search for "anomalies" from neutrino and anti-neutrino oscillations at large mass differences ($\Delta m^2 \approx 1eV^2$) with two LAr–TPC imaging detectors at different distances from the CERN-PS"*, SPSC-P-345 (2011).

[3] P. Bernardini et al., *"Prospect for Charge Current Neutrino Interactions Measurements at the CERN-PS"*, SPSC-P-343 (2011).

[4] G. Battistoni et al., *"The FLUKA code: Description and benchmarking"*, AIP Conf. Proc. 896, 31-49, (2007); G. Battistoni et al., *"A neutrino-nucleon interaction generator for the FLUKA Monte Carlo code"*, Proceedings of the 12th International conference on nuclear reaction mechanisms, Varenna (Italy), June 15 - 19, 2009, p.307; Ferrari et al., *"FLUKA, a multi particle transport code (program version 2005)"*, CERN-2005-10, INFN/TC-05/11, 2005.

[5] A. Aguilar et al. (LSND Collaboration), Phys. Rev. D 64, 112007 (2001); A. A. Aguilar-Arevalo (MiniBooNE Collaboration), Phys. Rev. Lett. 102, 101802 (2009); A. A. Aguilar-Arevalo et al. (MiniBooNE Collaboration), arXiv:1007.1150; F.Mills, ICHEP 2010, Paris, France; R. Van de Water, Neutrino 2010, Athens, Greece; E.D. Zimmerman, PANIC 2011, Cambridge, U.S.A.

[6] G. Mention et al., arXiv:1101.2755v1 [hep-ex] and previous references therein; J. N. Abdurashitov et al. (SAGE Collaboration), Phys. Rev. C 80, 015807 (2009); J. N. Abdurashitov et al. (SAGE Collaboration), Phys. Rev. Lett. 77, 4708 (1996); J. N. Abdurashitov et al. (SAGE Collaboration), Phys. Rev. C 59, 2246 (1999); J. N. Abdurashitov et al., Phys. Rev. C 73, 045805 (2006); F. Kaether, W. Hampel, G. Heusser, J. Kiko, and T. Kirsten, Phys. Lett. B 685, 47 (2010); P. Anselmann et al. (GALLEX Collaboration), Phys. Lett. B 342, 440 (1995); W. Hampel et al. (GALLEX Collaboration), Phys. Lett. B 420, 114 (1998).

[7] Some relevant papers are the following: J. Kopp, M. Maltoni, and T. Schwetz, "Are there Sterile-Neutrinos at the eV scale?", arXiv:1103.4570, (2011); Giunti, C. and Laveder, M., "3+1 and 3+2 Sterile Neutrino Fits", arXiv:1107.1452 (hep-ph), 2011.

[8] C. Rubbia, CERN-EP/77-08 (1977); ICARUS Coll., *"ICARUS initial physics program"*, ICARUS-TM/2001-03 LNGS P28/01 LNGS-EXP 13/89 add.1/01; ICARUS Coll., *"Cloning of T600 modules to reach the design sensitive mass"*, ICARUS-TM/2001-08 LNGS-EXP 13/89 add.2/01; C. Rubbia et al., JINST 6 P07011 (2011).

[9] www.gtt.fr, www.ihi.co.jp/en

[10] B. Baibussinov et at., JINST 5 (2010) P03005.

[11] S. Amerio et al., Nucl. Instr. And Meth. A527 (2004) 329.

[12]  M. Prata et al., Nucl. Instr. And Meth. A567 (2006) 222.

[13]  A. Ankowski et al., Nucl. Instr. And Meth. A556 (2006) 146.

[14]  V. Alvarez et al., "*SIPMs coated with TPB: coating protocol and characteriza-tion for NEXT*",  arXiv:1201.2019v1, 10 Jan 2012.

[15]  L. Bugel, J.M. Conrad, C. Ignarra, B.J.P. Jones, T. Katori, T. Smidt and H. K. Tanaka, "*Demonstration of a Lightguide Detector for Liquid Argon TPCs*", arXiv:1101.3013v1, 15 Jan 2011.

[16]  C. Carpanese et al., IEEE Trans. on Nucl. Science. Vol. 45 no. 4 (1998), 1804; C. Carpanese et al, Nucl. Instr. and Meth A409 (1998) 229; S. Centro et al., Nucl. Instr. and Meth A412 (1998) 440.

[17]  ATLAS Collaboration, "*ATLAS muon spectrometer: Technical Design Report*", ATLAS-TDR-010, CERN-LHCC-97-022.

[18]  CMS Collaboration, "*The CMS muon project: Technical Design Report*", CMS-TDR-003, CERN-LHCC-97-032.

[19]  ALICE Collaboration, "*ALICE dimuon forward spectrometer: Technical Design Report*", ALICE-TDR-5, CERN-LHCC-99-022.

[20]  S. Dusini et al., Nucl. Instr. and Meth. A508 (2003) 175.

[21]  ARGO collaboration, "*Astroparticle Physics with ARGO*" (1996, Proposal).

[22]  A. Paoloni et al., Nucl. Instr. and Meth. A583 (2007), 264.

[23]  S. Dusini et al., Nucl. Instr. And Meth. A602 (2009), 631.

[24]  A. Paoloni et al., Nucl. Instr. and Meth. A602 (2009), 635.

[25]  A. Paoloni et al., Nucl. Instr. and Meth. A661 (2012), S60.

[26]  M. Bazzi et al., Nucl. Instr. and Meth. A580 (2007) 1441.

[27]  R. Santonico, R. Cardarelli, Nucl. Instr. and Meth., A187, 377 (1981); R. Santonico, R. Cardarelli, Nucl. Instr. and Meth., A263, 20 (1988).

[28]  E. Ceron Zeballos et al., Nucl. Instr. and Meth., A392, 150 (1997).

[29]  G. Aielli et al. (ARGO-YBJ Collaboration), Nucl. Instr. and Meth. A661 S56 (2012).

[30]  D. Autiero et al., "*Design and Prototype Tests of the RPC system for the OPERA spectrometers*", RPC 2001 Workshop, Coimbra, November 2001.

[31]  R. Arnaldi et al., Nucl. Instr. and Meth. A490 51 (2002).

[32]  R. Acquafredda et al., JINST 4 (2009) P04018.

[33]  http://www.ohwr.org/projects/white-rabbit.

[34]  www.infn.it/CCR/server.

[35]  B. Baibussinov et at., JINST 5 (2010) P12006.

[36]  SAPOCO 42, EDMS 359387, November 2006.
      http://safety-commission.web.cern.ch/safety-commission/sapoco42/

[37] EN1993 Eurocode 3: Design of steel structures.
http://www.eurocodes.co.uk/eurocodes.aspx

[38] AISC, http://www.aisc.org/content.aspx?id=2884